\documentclass{article}
\usepackage{float}
\usepackage{ragged2e}
\usepackage{threeparttable}
\usepackage[T1]{fontenc}
\usepackage[utf8]{inputenc}
\usepackage[english]{babel}
\newenvironment{keyword}{\noindent\textbf{Keywords:} }{}
\usepackage[colorlinks=true, allcolors=blue]{hyperref}

\usepackage{newunicodechar}
\newunicodechar{；}{;}
\newunicodechar{，}{,}
\newcommand{\email}[1]{\href{mailto:#1}{\tt{\nolinkurl{#1}}}}
\newcommand{\orcid}[1]{ORCID: \href{https://orcid.org/#1}{\tt{\nolinkurl{#1}}}}

\usepackage[sfdefault,lf]{carlito}
\usepackage[parfill]{parskip}

\usepackage{fancyhdr}
\usepackage{placeins}
\usepackage{tabularx}
\usepackage{natbib}
\usepackage{authblk}
\usepackage{caption}
\usepackage{rotating}

\usepackage[a4paper,top=3cm,bottom=2cm,left=3cm,right=3cm,marginparwidth=1.75cm]{geometry}
\usepackage{amssymb}
\usepackage{pdfpages}

\usepackage{multirow}
\usepackage{adjustbox}

\usepackage{amsmath}
\usepackage{graphicx}
\usepackage{booktabs}
\usepackage{indentfirst}
\usepackage{orcidlink}

\usepackage[colorinlistoftodos]{todonotes}

\title{Assessing the Feasibility of a Video-Based Conversational Chatbot Survey for Measuring Perceived Cycling Safety: 
A Pilot Study in New York City}
\author[1,5]{Feiyang Ren}
\author[1,3，*]{Zhaoxi Zhang}
\author[2,4]{Tamir Mendel}
\author[1,2]{Takahiro Yabe}

\affil[1]{Center for Urban Science + Progress, Tandon School of Engineering, New York University, Brooklyn, 11201, USA}
\affil[2]{Department of Technology of Management and Innovation, Tandon School of Engineering, New York University, Brooklyn, 11201, United States of America}
\affil[3]{Department of Urban and Regional Planning, College of Design, Construction and Planning, University of Florida}
\affil[4]{School of Information Systems，The Academic College of Tel Aviv-Yaffo， Tel Aviv，Israel}
\affil[5]{School of Geography, University of Leeds, UK}

\affil[*]{Corresponding author:  \email{zhang.zhaoxi@ufl.edu}}
\date{}  

\begin{document}
\maketitle
\thispagestyle{fancy}

\begin{abstract}

Bicycle safety is important for bikeability and transportation efficiency. However, conventional surveys often fall short in capturing how people actually perceive cycling environments because they rely heavily on respondents' recall rather than in-the-moment experience. By leveraging large language models (LLMs), this study proposes a new method of combining video-based surveys with a conversational AI chatbot to collect human perceptions of cycling safety and the reasons behind these perceptions. The paper developed the AI chatbot using a modular LLM architecture, integrating prompt engineering, state management, and rule-based control to support the structure of human-AI interaction. This paper evaluates the feasibility of the proposed video-based conversational chatbot using complete responses from sixteen participants to the pilot survey across nine street segments in New York City. The method feasibility was assessed using a seven-point scale rating for user experience (i.e., ease of use, supportiveness, efficiency) and a five-point scale for chatbot usability (i.e., personality, roboticness, friendliness), yielding positive results with mean scores of 5.00 out of 7 (standard deviation = 1.6) and 3.47 out of 5 (standard deviation = 0.43), respectively. The data feasibility was assessed using multiple techniques: (1) Natural language processing (NLP), such as KeyBERT, for overall safety and feature analysis to extract built-environment attributes; (2) K-means clustering for semantic analysis to identify reasons and suggestions; and (3) regression to estimate the effects of built-environment and demographic variables on perceived safety outcomes. The results show the potential of AI chatbots as a novel approach to collecting data on human perception, behavior, and future visions for transport planning.



\end{abstract}

\section{Introduction}
Bicycling is recognized as a sustainable and health-promoting form of active transportation that reduces traffic congestion and vehicle emissions \citep{safety9040075,JEON2024103992}. Existing literature shows that the built environment (i.e., street crossings \citep{BI2023103551}) influences cyclists' behavior \citep{JEON2024103992}, risk perception \citep{KWON2020105716,JEON2024103992}, and travel patterns \citep{BI202251}.  These dimensions are closely tied to perceived cycling safety \citep{JEON2024103992}, and concerns about safety present a major barrier to increasing the share of cycling as a mode of urban travel \citep{LAWSON2013499}. Bicycle traffic is also threatened by accidents involving vulnerable road users, such as cyclists, moped riders, and motorcyclists, who are not protected by a vehicle body and therefore face a heightened risk of severe injury or death \citep{Skoczynski2021, HE2026104115}. The provision of safe and accessible cycling infrastructure encourages bicycle use, whereas inadequate facilities deter cycling and increase the risk of severe or fatal injuries \citep{JEON2024103992}. While existing studies have primarily focused on the objective measurement of cycling infrastructure by using observational and GIS-based methods to quantify built environment attributes \citep{Ma2014,BLITZ202127}, these measures may not fully align with individuals' perceptions of the cycling environment. In previous work, personal perceptions of local environmental characteristics can differ significantly from objectively measured features \citep{BLITZ202127}. As perceived measures reflect how individuals interpret and experience the built environment, discrepancies often emerge between objective and perceived measures \citep{Ma2014}. Additionally, previous studies have emphasized that people's responses to survey or interview questions about perception are largely based on their recall \citep{althubaiti2016information}. Addressing this disconnect is essential for understanding cycling behavior and enhancing safety outcomes. With the emerging opportunities of new technology, an increasing number of researchers are investigating how to improve the methodology for measuring human perception of the environment such as multi-sensor information fusion \citep{wevj16010020}, immersive environments enabled by Virtual Reality (VR)\citep{VR2025Al, zhangUsingVirtualReality2026} and interactive conversational AI chatbots \citep{zhang2025human}. This paper adopts an innovative approach involving an AI chatbot with videos to explore the feasibility of using it to study public perceptions of cycling safety. Furthermore, by evaluating participant feedback on human-chatbot interaction and data potential, this paper investigates the value of video-based AI chatbots in collecting human feedback to qualify human perception in urban studies. 


\section{Literature Review}

\subsection{Perceived Cycling Safety}
In the existing literature, cycling represents a critical solution to the persistent challenges of traffic congestion, air quality, and public health \citep{PUCHER2010S106, Lusk2013, nieuwenhuijsen2020urban, DAI2021101013}. However, the realization of these benefits is significantly compromised by the complex and multidimensional nature of perceived cycling safety, which has been identified as a key determinant of cycling behavior \citep{parkinChapter6Network2012, MANTON2016138, safety9040075}. Perceived safety influences not only current cyclists but also prospective mode users\citep{parkinModelsPerceivedCycling2007}, applying multiple scales to the decision to cycle, individual route choice, and actual behavior on the road \citep{MANTON2016138, moran2025causal}.

Perceived cycling safety is a multidimensional construct that reflects an individual’s interaction with the actual environment, involving awareness and perception of the outside world through their primary receptive senses \citep{Ma2014}. Empirical evidence highlights the severity of these perceived risks. \citet{MARSHALL2019types} noted that fatality and nonfatal injury rates for bicyclists are nearly twice the overall average for all transport modes. In addition to traffic risks, perceived cycling safety encompasses social security, particularly the risk of becoming a victim of crime, such as theft or personal assault, which further complicates the provision of a safe operating environment \citep{ceccato2014safety, marquezIntegratingPerceptionsSafety2021}. Moreover, recent research has extended the components of perceived cycling safety to include contaminant exposure, street conditions, and weather conditions \citep{safety9040075}.

To conclude, cycling safety is influenced by a combination of subjective perceptions and objective environmental conditions \citep{Ma2014}. The consideration of these perceptions is central to the successful cycling planning and design \citep{parkinChapter6Network2012}. Although numerous studies have highlighted the significance of human perception regarding cycling safety， particularly in relation to road quality and infrastructure design, current measurement methodologies remain inadequate for capturing the multidimensional and situational factors that shape these perceptions \citep{safety9040075}. Traditional surveys and questionnaires have been widely applied to examine human's psychological responses to urban topics (i.e. the effects of urban nature \citep{AIApproach2019}), but they are inherently static \citep{safety9040075} and rely on participants' memories \citep{althubaiti2016information}. Consequently, such conventional methods might struggle to capture the dynamic interplay of infrastructure, traffic behaviors, and perceptions during an active journey \citep{BLITZ202127, Yu2022Exploring}. Therefore, there is a critical gap between the conceptual understanding of perceived cycling safety and the ability to measure it across diverse, real-world urban contexts \citep{ZENG2024103739}.

 
\subsection{Urban Features Related to Cycling Safety}

The built environment affects individual behaviors through its impact on perception during one's active engagement with the surroundings \citep{Ma2014}. Consequently, well-designed landscapes should be considered in any development to enhance active travel and recreational cycling behavior among the residents \citep{ETMINANIGHASRODASHTI2018241}. Table~\ref{tab:feature_summary} summarizes the features of the built environment identified in the literature. Drawing on the framework by \citet{BLITZ202127}, these environmental influences can be categorized into dimensions such as cycling infrastructure, traffic environment, and public space quality.


Empirical research highlighted that physical infrastructure is a critical determinant of safety outcomes. Protected or delineated bicycle lanes and cycle tracks support more predictable cyclist movements and reduce conflicts and injury risks, leading to corridor level interventions \citep{huang1999comparative, hunter2000evaluation, MINIKEL2012241}. While conventional painted lanes offer more modest benefits primarily by improving lateral positioning, they failed to eliminate exposure to potential traffic conflicts. In contrast, shared lanes show no significant safety improvements and often fail to reduce crash risk \citep{MARSHALL2019types}. Beyond lane types, roadway geometry, including lane width, intersection geometry, sidewalk design, and quality, influences both perceived safety and the actual likelihood of crashes \citep{DAI2021101013,PUCHER2010S106}. For instance, surface roughness and steep gradients affect stability and braking distances \citep{PUCHER2010S106, Teschke2012slope}. 

Moreover, the traffic environment and its spatial and temporal interactions of cycling with other modes represent another critical dimension. For example, on-street parking elevates the risk of dooring—being hit by a suddenly opened car door \citep{SCHIMEK2018sideparking}, while interactions with motorized vehicles increase psychological uncertainty for cyclists \citep{Ivan2023motor}. Cyclists also face mobility impacts from interactions with pedestrians and other non-motorized users \citep{Dozza2014cyclistpedestrain}. However, a higher volume of cyclists can promote a "safety in numbers" effect, where motorists adapt their behavior by slowing down and providing more space, while groups of cyclists collectively detect and signal potential hazards \citep{Jacobsen2003SafetyInNumbers}.

Finally, public space quality, including streetscape greenery and vegetation buffers, influences comfort and perceived safety by providing visual separation from traffic and enhancing aesthetic quality \citep{Ricchetti2025Greenery}. However, poorly maintained vegetation can limit sight distances at crossings, increasing risk \citep{Yu2024StreetGreenery}. Trip purpose further moderates these environmental effects; commuters and recreational riders often exhibit varying tolerances for environmental stressors and choose different routes based on their specific needs \citep{LAWSON2013499}. Thus, the interaction between physical features and user intent is fundamental to the conceptualization of a safe cycling environment.

While Table~\ref{tab:feature_summary} provides a comprehensive catalog of the physical determinants of cycling safety, a significant methodological gap persists. Existing research often relies on static, GIS-based physical metrics in isolation, which frequently fails to account for the synergistic and dynamic effects of these features as perceived by cyclists in motion \citep{Yu2022Exploring, Ewing01022009}. To bridge this gap, research requires methodologies that move beyond passive observation to capture active, real-time interpretations of the environment.

\begin{table}[H]
\centering
\caption{Built Environment Features Influencing Cycling Safety}
\label{tab:feature_summary}
\begin{tabular}{p{4cm} p{5cm} p{5cm}}
\toprule
\textbf{Feature} & \textbf{Key Concepts} & \textbf{Citation} \\
\midrule

Bike Lane Type & Modal separation and crash likelihood vary by design & \cite{MARSHALL2019types} \\
Bike Lane Width & Space, comfort, lateral clearance & \cite{DAI2021101013} \\

Car Lane Width & Speed reduction, vehicle behavior & \cite{DAI2021101013} \\

Road Surface & Roughness, debris, stability & \cite{PUCHER2010S106} \\

Greenery & Separation, visual quality, sight distance risk & \cite{Yu2024StreetGreenery,Ricchetti2025Greenery} \\

Road Grade & Speed, braking, downhill risk & \cite{Teschke2012slope} \\

Construction & Obstruction, forced merging & \cite{huang1999comparative} \\

Sidewalk & Modal separation & \cite{DAI2021101013} \\

Side Parking & Collisions with Car doors & \cite{SCHIMEK2018sideparking} \\

Crossings & Conflict points, turning risk & \cite{DAI2021101013} \\

Motorcyclists & Short passing distance, unpredictability & \cite{Ivan2023motor} \\

Pedestrians & Shared path conflicts & \cite{Dozza2014cyclistpedestrain} \\

Cyclists & Overtaking, weaving, density effects; safety in numbers, driver awareness, collective vigilance & \cite{Dozza2014cyclistpedestrain,Jacobsen2003SafetyInNumbers} \\

Car Volume & Crash severity, speed differential & \cite{JEON2024103992} \\

\bottomrule
\end{tabular}
\end{table}

\subsection{Chatbot Technology}
\label{sec:chatbottech}

In the past decade, many cities have already adopted rule-based chatbots in various industrial sectors,  including customer service, e-commerce, healthcare, banking and finance, travel and hospitality, human resources, education, entertainment, government, and real estate, among others \citep{Senadheera01022024}. Their implementation reduces the workload for local government staff \citep{ANDROUTSOPOULOU2019358,Henman01102020,Senadheera01022024}, while streamlining the processing and response to citizen inquiries \citep{Senadheera01022024,Henman01102020}. In addition to the top-down administrative functions, chatbots have been increasingly recognized for their potential in fostering bottom-up participation, enabling citizens to actively contribute localized knowledge and lived experiences to the planning process \citep{falcoDigitalParticipatoryPlatforms2018, afzalanRoleSocialMedia2014}. Such conversational interfaces facilitate a more responsive feedback loop \citep{Hasan-2023}. This dual role positions chatbots not only as efficient service providers but also as digital intermediaries that facilitate more inclusive urban governance \citep{land10010033}.

While rule-based systems provide a structured foundation for these roles, the emergence of Generative AI chatbots amplifies the bottom-up capacity of urban planning \citep{AIApproach2019,Senadheera01022024}. Generative AI chatbots are computer programs capable of understanding user inquiries and responding automatically using large language models (LLMs), machine learning (ML), dialogue management, application programming interfaces (API), and natural language generation (NLG) \citep{Senadheera01022024}. Unlike traditional static systems, these generative interfaces are designed to interact in a human-like manner, providing citizens with up-to-date information and actionable insights \citep{Hasan-2023}. Specifically, \cite{zhang2025human} compared the AI chatbot with a traditional static one, finding that the former significantly improves the depth of user engagement and the richness of collected data. The interactive methodologies are increasingly vital for moving beyond foundational roles in transportation and emergency management, to address broader structural crises, ranging from local traffic congestion to global climate change \citep{Hasan-2023, Peng2024}. With the continued advancement of AI technologies, there is growing anticipation for the development of innovative and efficient chatbot applications to enhance the effectiveness and efficiency of urban planning processes \citep{HERATH2022100076, Hasan-2023}.



To address the methodological gaps identified in the preceding sections, particularly the limitations of static spatial metrics and the inherent biases in traditional subjective surveys, this study proposes an integrated multimodal framework. Building on the evolution of AI and LLM, the study aims to develop a conversational AI chatbot that integrates videos to capture multiple dimensions of perceived cycling safety and related built environment features in motion. The integration of these videos serves as a critical visual anchor that mitigates perception biases (i.e. response and recall bias). Furthermore, this approach controls for confounding factors by providing immersive and real-time stimuli during the data collection process \citep{althubaiti2016information, itoUnderstandingUrbanPerception2024}. By facilitating a real time dialogue between participants and the AI chatbot, the study moves beyond the limitations of static, human memory-based surveys and aims to understand and measure perceived safety across diverse urban contexts.

\section{Research Goals}
Using New York City as an example, this paper aims to develop a video-integrated AI chatbot survey that enables users to perceive cycling safety from a first-person perspective in real-world videos. To evaluate the feasibility and insights of this approach, the study focuses on three objectives: 1) Design and develop a controllable conversational workflow for an AI chatbot embedded in a video-based survey, enabling consistent task completion across comparable scenarios; 2) Evaluate both method and data feasibility of the chatbot-based survey methodically through user testing; 3) Assess the feasibility of using the collected text data from human-chatbot conversations to assess human perceptions of cycling safety.   

By implementing a pilot study in New York City, the research aims to provide new insights into eliciting granular, situational feedback, moving beyond static ratings to uncover the underlying semantic reasoning of cyclists. Furthermore, this study contributes a standardized workflow and a replicable methodology for AI-driven urban data collection, fostering a more human-centric understanding of transportation safety that informs evidence based infrastructure design for vulnerable road users.

\section{Chatbot-based Survey Design}

\label{sec:Methodology}
\subsection{Selection of Cycling Roads}
To situate and test our video-based AI chatbot technology, this study used New York City (NYC) as a representative case study. NYC provides an ideal testing ground due to its rapidly expanding cycling culture and diverse urban environments. According to the New York City Department of Transportation (DOT), the city sees an estimated 620,000 daily cycling trips, with over 762,000 residents cycling regularly \citep{nycDOTCycling}. This high volume of active users, combined with the city's complex streetscapes, offers a robust context for evaluating how conversational AI can capture safety perceptions across a wide demographic of urban cyclists.

\begin{figure}[h!]
    \centering
    \includegraphics[height=12cm,keepaspectratio]{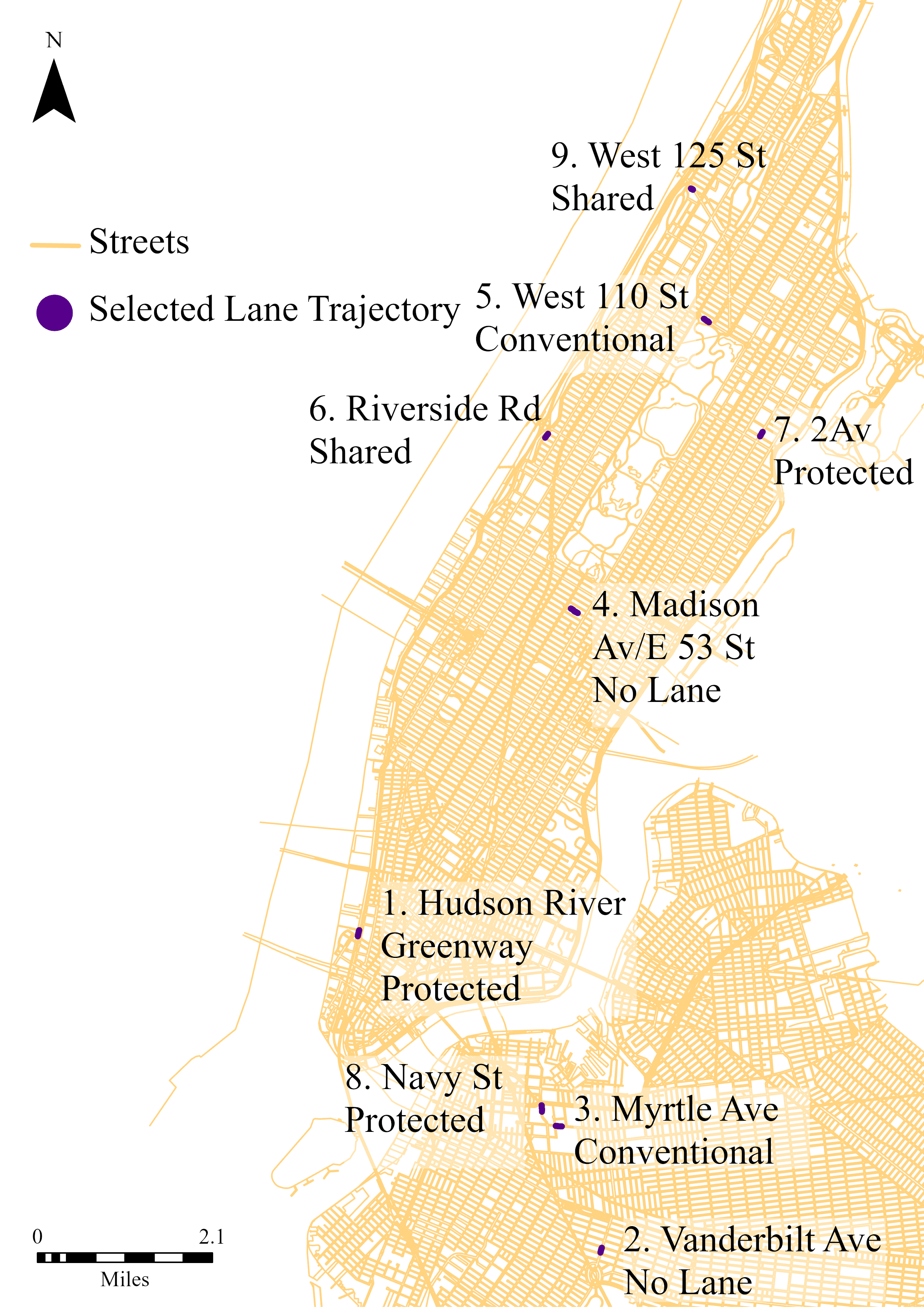}
    \caption{Spatial overview and street level representations of the nine numbered selected NYC segments, the road data was adapted from \cite{NYCDCPDigitalMap}. All participants viewed the nine videos in an identical sequence as the street numbers on the map, ensuring consistency in exposure and eliminating order variation across users. A total of nine street segments were selected through purposive sampling to ensure structural variation in both cycling infrastructure and urban morphology. Selection criteria included bike lane types (protected, conventional, shared, and no dedicated lane), as well as representation across NYC’s horizontal and vertical grid structure, capturing avenues, cross streets, and waterfront corridors.}
    \label{fig:streetmap}
\end{figure}

As illustrated in Figure \ref{fig:streetmap}, the street network in New York City follows a grid pattern \citep{tian2025quantifying} primarily characterized by horizontal streets (running east-west) and vertical streets (running north-south), as designed by the Commissioners' Plan of 1811. The guidelines \citep{NYCStreetDesignBikeLane2020}: 1) A shared lane, where bicycles and motor vehicles coexist within the same space, typically indicated by pavement markings to alert drivers to the presence of cyclists; 2) A conventional bike lane, a designated cycling space marked by painted lines, providing cyclists with a dedicated area adjacent to vehicular traffic; 3) A protected bike lane, which enhances safety by incorporating physical barriers such as bollards, curbs, or parked vehicles to separate cyclists from moving traffic \citep{NYCStreetDesignBikeLane2020}. 

Considering the various types of bike lanes, the researchers selected nine segments in Manhattan and Brooklyn for video collection, as illustrated in Figure \ref{fig:streetmap}. The selected segments are as follows: 1) Hudson River Greenway, a protected bike lane in Lower Manhattan; 2) Vanderbilt Avenue, a street without a bike lane in the Park Slope neighborhood of Brooklyn; 3) Myrtle Avenue, a standard bike lane in Downtown Brooklyn; 4) Madison Av and East 53rd Street in Midtown Manhattan, which lacks a designated bike lane; 5) West 110th Street, featuring a conventional bike lane in the Upper West Side; 6) Riverside Drive, a shared bike lane in the Upper West Side; 7) 2nd Avenue, a protected bike lane in the Upper East Side; 8) Navy Street, a protected bike lane in Downtown Brooklyn; 9) West 125th Street, a shared bike lane in Harlem. All videos were recorded in situ during actual rides along the selected roads. A camera securely mounted beneath the bicycle’s handlebar was used to capture the first-perspective view of the cycling environment.

\subsection{Feature Evaluation}
\label{objectfeatureevaluation}

From the existing literature, a total of thirteen built environment features were selected, as shown in Table \ref{tab:feature_eva}. Three experts with diverse backgrounds in urban planning and cycling were invited to independently assess the selected features using expert based audit instruments \citep{Brownson2009, Ewing01022009}. Binary features, including slope, construction, sidewalk, side parking, crossing, motorcyclists, pedestrians, and other cyclists, were coded as 0 if they were "absent/non existing" or 1 if they were "present/existing". Other features including bike lane width, car lane width, road surface quality, greenery, and car presence, were rated on a four-point ordinal scale (0–3) where 0 represents non existence, 1 is low, 2 is medium, and 3 is high. Although the scale is ordinal by design, intermediate values (e.g., 1.5) were permitted to reflect nuanced expert judgment, and the resulting scores were treated as continuous in subsequent analysis. For each section $i$ and feature $f$, the raw expert ratings $r_{if}^{(e)}$ were averaged across the three experts and then rescaled to the unit interval to obtain a normalized feature score $\tilde{s}_{if} \in [0,1]$. The scoring rubric assumes that every expert rated variable contributes equally to the overall measure of that feature. The resulting mean value provides a single representative indicator for the feature, making it possible to compare consistently across different street segments. In parallel, categorical attributes such as road classification or surrounding land use were kept in their original form to preserve contextual information. These scores and attributes serve as the baseline for the survey's backend in \ref{subsec:chatbottechnical}.

\begin{table}[!ht]
    \centering
    \begin{threeparttable}
    \caption{Expert Evaluation of Built Environment Features across Selected NYC Street Segments.} 
    \label{tab:feature_eva}
    \renewcommand{\arraystretch}{1.2} 
    \setlength{\tabcolsep}{3pt} 

    \begin{tabularx}{1.0\textwidth}{X|ccccccccc} 
        \toprule
        \textbf{Attribute} & \textbf{Hudson} & \textbf{Vander-} & \textbf{Myrtle} & \textbf{Madison} & \textbf{West} & \textbf{River-} & \textbf{2 Av} & \textbf{Navy} & \textbf{West} \\
        & \textbf{Greenway} & \textbf{bilt} & \textbf{Ave} & \textbf{Av} & \textbf{110 St} & \textbf{side} & & \textbf{St} & \textbf{125 St} \\
        \midrule
        Lane Type & Protected & Mixture & Conv. & None & Conv. & Shared & Prot. & Prot. & Shared \\
        \midrule
        Bike Lane Exist (B) & Y & Mixture & Y & N & Y & Y & Y & Y & Y \\
        Bike Lane Width (O)  & 1.000 & 0.000 & 0.444 & 0.000 & 0.444 & 0.278 & 0.556 & 0.611 & 0.000 \\
        Car Lane Width (O)   & 0.000 & 0.778 & 0.889 & 0.667 & 1.000 & 1.000 & 1.000 & 0.889 & 1.000 \\
        Road Surface (O)     & 1.000 & 0.000 & 1.000 & 0.333 & 0.667 & 0.444 & 0.444 & 1.000 & 0.333 \\
        Greenery (O)       & 1.000 & 0.333 & 0.444 & 0.111 & 0.222 & 0.222 & 0.111 & 0.111 & 0.000 \\
        Road Grade (B)     & 0.333 & 0.667 & 0.667 & 0.333 & 0.667 & 0.667 & 0.667 & 0.667 & 0.000 \\
        Construction (B)     & 0.0 & 1.0 & 0.0 & 0.0 & 1.0 & 0.0 & 1.0 & 0.0 & 1.0 \\
        Sidewalk (B)      & 1.0 & 1.0 & 1.0 & 1.0 & 1.0 & 1.0 & 1.0 & 1.0 & 1.0 \\
        Side Parking (B)    & 0.000 & 0.333 & 1.000 & 1.000 & 1.000 & 1.000 & 1.000 & 0.000 & 1.000 \\
        Crossings (B)         & 0.667 & 1.000 & 0.333 & 1.000 & 0.000 & 0.667 & 1.000 & 0.000 & 0.000 \\
        Motorcyclists (B)    & 0.0 & 1.0 & 0.5 & 0.0 & 0.0 & 0.0 & 0.0 & 0.0 & 0.0 \\
        Pedestrians (B)       & 1.000 & 1.000 & 1.000 & 1.000 & 0.333 & 0.000 & 1.000 & 0.000 & 0.000 \\
        Cyclists (B)          & 1.0 & 1.0 & 0.0 & 0.0 & 0.0 & 0.0 & 1.0 & 1.0 & 0.0 \\
        Car Volume (O)              & 0.000 & 0.778 & 0.333 & 1.000 & 0.333 & 0.222 & 0.111 & 0.000 & 0.778 \\
        \bottomrule
    \end{tabularx}
    
    \begin{tablenotes}
            \footnotesize
            \item Expert-based normalized assessment of built-environment attributes across nine selected NYC street segments. For each segment, three independent evaluators rated infrastructural design (e.g., bike lane width, road surface), environmental context (e.g., greenery, road grade), and dynamic traffic conditions (e.g., car volume, cyclists). Binary attributes (B) are coded as 0–1 indicators for existence/non-existence, while ordinal variables (O) derive from a four-point (0-3) ordinal scale. Ratings were averaged and linearly rescaled to the [0,1] interval to ensure comparability across heterogeneous feature types. The resulting normalized scores serve as baseline feature weights within the chatbot system, anchoring segment specific responses after participants select particular features. 
.
        \end{tablenotes}
    \end{threeparttable}
\end{table}

\subsection{Chatbot-based Survey Design}
\label{subsec:surveydesign}
\begin{figure}[p]
    \centering
        \includegraphics[width=0.9\textwidth]{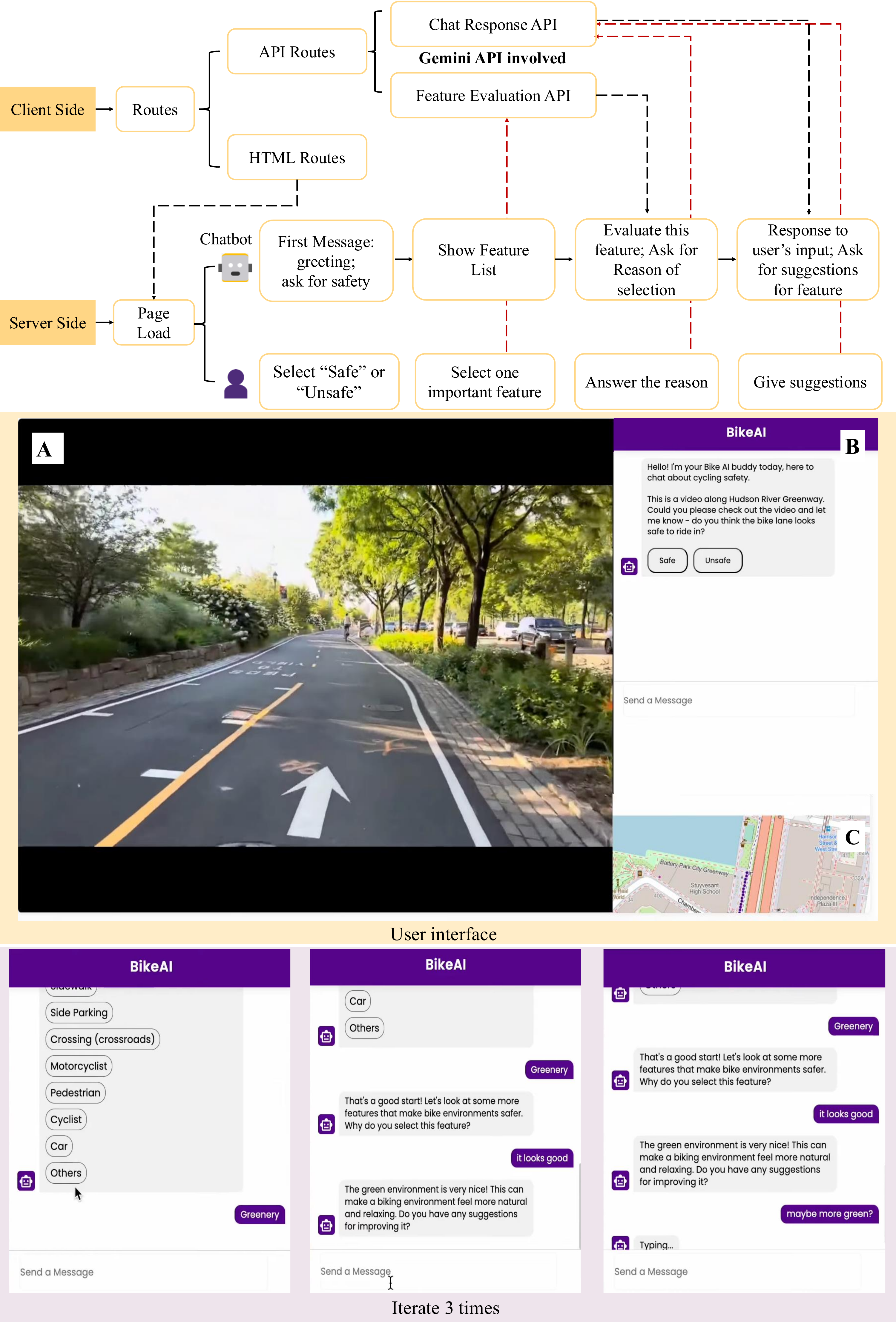}
    \caption{Chatbot design and user interface. Participants viewed short first-person cycling videos recorded across different types of bicycle facilities in New York City and interacted with an AI chatbot to report their overall safety perceptions ("safe"/"unsafe") and related built-environment features included in Table \ref{tab:feature_summary}. Each interaction generated structured conversation records, including timestamps and anonymized user identifiers, enabling responses to be linked across video segments while preserving participant anonymity. The user interface consists of a video panel (A), a chatbot window (B), and a map showing the corresponding riding segment (C). Black dashed lines represent client–server request–response communication (HTML and API routes). Red dashed lines indicate server-side API invocations to the Gemini model for generating conversational responses and feature evaluations.}
    \label{fig:design_and_interface}
\end{figure}

The flowchart in Figure \ref{fig:design_and_interface} outlines a client-server architecture for the interactive, video-integrated chatbot that evaluates perceived cycling safety features. The system includes a server side application built with Flask, featuring API and HTML routes, as well as a user interface that utilizes JavaScript to facilitate user interactions. The user interface includes a video clip (A) to present road conditions, a chatbot interface (B) for human-chatbot interaction, and an embedded map (C) indicating the specific segment being evaluated. The AI chatbot was built based on a large language model (LLM), Gemini [Google] (specifically, the gemini-2.0-flash-001), and deployed on the Google Cloud Platform with a temperature of 1.0, top-p of 0.95, and a maximum output of 8192 tokens.

The interaction process is designed as a state-driven dialogue that guides users through a structured evaluation in four sequential steps shown in Figure \ref{fig:design_and_interface}:

\begin{itemize}
    \item \textbf{Step 1: Initial Perception} The user first watch a first-person perspective cycling video that simulates the real-world riding experience. As the video plays, the AI chatbot introduces the specific road segment being shown, situating the user in the environment. The participant then provides an immediate, holistic assessment of the scene by responding to the question: \textit{"do you think the bike lane looks safe to ride in?"}
    \item \textbf{Step 2: Cycling Safety Perceptions towards Features} After making a safety judgment, the user browses a list of the built environment features (i.e. bike lane width, greenery, side parking) to select the factor that most significantly influenced their decision. Instead of displaying raw scores, the AI chatbot internally processes the technical condition of the feature as mapped from expert evaluation scores from Table \ref{tab:feature_eva}. The system then translates these metrics into a concise and natural language statement. Consequently, the user receives an AI-generated comment reflecting the infrastructure quality or traffic intensity specific to that segment rather than seeing a numerical value or a simple label.
    \item \textbf{Step 3: Articulating Subjective Reasoning} Upon seeing the AI’s descriptive statement, the user encounters the inquiry: \textit{"Why do you select this feature?"} At this stage, the user provides their own reasons, which gives them the opportunity to either elaborate on the AI's points or offer a different perspective based on what they personally observe in the video.
    \item \textbf{Step 4: Proposing Improvements and Iteration} Finally, the user receives an empathetic response and addresses the follow-up question: \textit{"Do you have any suggestions for improving it?"} This allows the participant to act as a co-designer by suggesting practical interventions for the cycling environment.
\end{itemize}

\subsection{Chatbot Technical Implementation}
\label{subsec:chatbottechnical}

To ask questions, receive participants' feedback, and ensure quality and consistency, the initial context sent to the LLM included instructions about the study objective, the role of the chatbot, and the expected interaction style. The prompts were framed as: \textit{“Answer as if you are helping users identify the features of a safe or unsafe bike environment based on the provided video. Respond empathetically; be verbal, informal, interesting, and friendly. Every response MUST be less than 15 words. Ask one question at a time and avoid repetition. Do not use phrases such as ‘Hey there’ or ‘How are you?’ and do not pose questions back to the participant.”}

The technical backend operates through four sequential steps that correspond to the participant interaction flow:

\begin{itemize}
    \item \textbf{Step 1: Session Initialization} The AI chatbot first introduces itself and displays the name of the road segment shown in the current video. The system then records the participant’s classification of the cycling environment as either "safe" or "unsafe".
    \item \textbf{Step 2: Backend Feature Processing and Evaluation API} Upon the selection of a feature, the system retrieves the predefined evaluation score from an internal baseline defined in Table \ref{tab:feature_eva}. For binary features such as the presence of construction barriers or side parking, the function maps positive scores to "exists" and zero scores to "does not exist". For continuous features including bike lane width, surface quality, or greenery, the system applies a three-tier categorization where scores above 0.67 are classified as "good", scores between 0.33 and 0.67 are "average", and scores between 0 and 0.33 are "poor", while a score of 0 indicates the feature does not exist. The "Feature Evaluation API" then triggers the first call to Google’s Gemini API, transmitting the feature name and derived condition to the server to generate a concise, segment specific evaluation statement.
    \item \textbf{Step 3: Conversational Reasoning via Chat Response API} The "Chat Response API" manages the conversational stage through a second and independent call to Google’s Gemini API. The server returns the AI-generated evaluation statement combined with the follow-up question, \textit{"Why do you select this feature?"} within a single dialog block. This configuration allows the system to receive the user's natural language reasoning and store the conversation history on the server, where Google's Gemini API generates a contextual comment based on the prior messages.
    \item \textbf{Step 4: Feedback Solicitation and Data Storage} In the final technical stage, the system displays the generated comment followed by the question: \textit{"Do you have any suggestions for improving it?"} to solicit improvement feedback. After the user provides suggestions, the system flow prompts the selection of the next important feature for a total of three iterations. All collected data is stored in the Google Cloud Platform for further analysis.
\end{itemize}

\section{Pilot Study}
We conducted an online pilot study between June 30 and October 8, 2025. In total, 41 responses were recorded, where a response was registered each time the survey was accessed. After excluding internal test entries and incomplete submissions, 16 unique participants (39.0\%) completed the full evaluation of all nine video segments, yielding 144 valid cycling safety evaluations. Because one participant did not complete the full demographic questionnaire, the mixed-effects logistic regression was estimated on 135 participant-section observations from the remaining 15 respondents. Participants were recruited through the university social space and social media. All procedures were reviewed and approved by the Institutional Review Board (IRB), and each participant received \$5 in compensation upon the successful completion of the survey. To ensure eligibility, participants were required to provide informed consent and confirm that they lived in New York City, were over 18 years of age, and were English speakers by answering “Yes” to the corresponding screening questions. Of the 16 responses received, a total of 15 participants completed the demographic survey. Among them, 53.3\% identified as male and 46.7\% as female. In terms of age, two-thirds of the sample (66.7\%) were between 18 and 24 years old, followed by 26.7\% between 25 and 34, and 6.7\% between 35 and 44 years old. Cycling frequency among participants was generally low. A large majority (86.7\%) cycled less than once per week, while only 6.7\% reported cycling daily and another 6.7\% cycled two to six times per week. Awareness of cycling related laws also varied considerably. Over half of the respondents (53.3\%) reported knowing only one cycling law, while 13.3\% knew two laws, 26.7\% knew three laws, and 6.7\% knew four laws. No respondents reported having no legal knowledge.

\section{Exploration of Feasibility}

This section evaluates the feasibility of the video-based AI chatbot technology through two lenses: method feasibility and data feasibility. For a newly emerging methodology, conducting a comprehensive feasibility study is indispensable to validate both the technical approach and user acceptance  \citep{zhangFeasibilityIntegratingWearable01}. Method feasibility assesses the functional performance and user-centered viability of the AI chatbot as the survey instrument. Furthermore, user-centered viability is evaluated through participant experience; as noted in recent literature, user-friendliness, ease of use, and specifically perceived naturalness of AI are critical factors for the practical success of new data collection tools \citep{zhangFeasibilityIntegratingWearable01,jacobsenChatbotsDataCollection2025}. In this context, method feasibility focuses on whether the chatbot's conversational architecture can effectively navigate the survey process.

While method feasibility confirms the operational viability of the interaction, data feasibility examines the utility of the captured information. This includes assessing the richness of participants' responses, their perceptions regarding cycling safety, and their reasons and suggestions. Beyond the content itself, the feasibility of digital data collection fundamentally depends on privacy protection and data security \citep{zhangFeasibilityIntegratingWearable01}. To address these requirements, all captured data are securely managed and stored within Qualtrics and Google Cloud Storage. Furthermore, during the data processing phase, all personally identifiable information and any metadata that could potentially track participant identity are removed during analysis. Together, these metrics validate the AI chatbot as a robust and ethical alternative to traditional urban data collection methods.

\subsection{User Test for Evaluations}
Two validated usability surveys were used to evaluate method feasibility. The Short User Experience Questionnaire (UEQ-S) was employed to measure the overall user experience by assessing pragmatic quality, such as efficiency and dependability, alongside hedonic quality, including stimulation and novelty \citep{schrepp2017design, ueqshort_items}. Additionally, the Chatbot Usability Questionnaire (CUQ) was used to evaluate human-computer interaction specific to conversational agents, focusing on the chatbot's personality, navigation ease, and dialogue flow \citep{holmesChatbotUsabilityQuestionnaire}. These instruments collectively provide a comprehensive assessment of how effectively the AI chatbot functions.

\subsection{Text Mining}
To extract safety-relevant information from user responses, we implemented a mixed NLP workflow for text mining. Before analysis, all user messages were cleaned (lowercasing, removal of obvious stopwords and system artifacts) and encoded using a sentence transformer model (all-MiniLM-L6-v2), which maps each phrase into a dense semantic embedding in a shared vector space. Two analyses have been conducted: 

1) Feature analysis: we applied KeyBERT \citep{grootendorst2020keybert} to each message to identify the most informative noun phrases and short expressions that capture how participants described specific features. These key phrases were then embedded using the same transformer model to ensure consistency between the sentence level and phrase level representations. The results are presented in Section \ref{subsec:Feature_Analysis}.

2) Semantic analysis: we performed K-means clustering on the resulting embedding matrix to group semantically similar phrases. The number of clusters was selected based on a combination of internal validity indices (e.g., silhouette trends) and interpretability, resulting in a small set of coherent thematic categories of perceived environmental features. This pipeline allows us to transition from raw, free text descriptions to structured semantic clusters that can be related back to lane types and safety labels in subsequent analyses. The results are presented in Section \ref{subsec:semantic}. 


\subsection{Mixed-effects Logistic Regression}
A mixed-effects logistic regression was used to examine the association between the built environment characteristics included in Table \ref{tab:feature_summary} and perceived cycling safety. In our analysis, each observation represents a unique evaluation of a video segment by a specific participant (identified by a User ID). The dependent variable indicates whether a video segment was perceived as safe. The model can be written as:
\begin{align}
  \text{safe}_{iv} \mid u_i, b_v &\sim \text{Bernoulli}(p_{iv}),\ \operatorname{logit}(p_{iv}) = \alpha + \mathbf{X}_{i}^{\top}\boldsymbol{\beta} + \mathbf{Z}_{iv}^{\top}\boldsymbol{\gamma} + u_i + b_v,
\end{align}
where $u_i \sim \mathcal{N}(0,\,\sigma_u^2)$ and 
$b_v \sim \mathcal{N}(0,\,\sigma_v^2)$ are independent random intercepts for participant $i$ and video section $v$, respectively. $\mathbf{X}_{i}$ is a vector of participant-level demographic controls (cycling frequency, gender, age group, education, and race) together with a behavioral count variable; $\mathbf{Z}_{iv}$ is a vector of expert-coded built environment features normalized to $[0,1]$. Categorical variables were dummy-coded with the most frequent category as the reference. Model parameters were estimated via maximum likelihood. 
To facilitate interpretation, average marginal effects (AMEs) were 
computed as the sample average of individual-level partial effects, expressing results as changes in predicted safety probability. The results are presented in Section~\ref{subsec:regression}.

\section{Results}

\subsection{Overall Safety}
\label{subsec:overallsafety}

\begin{figure}[h!]
    \centering    \includegraphics[width = 1.0\textwidth,keepaspectratio]{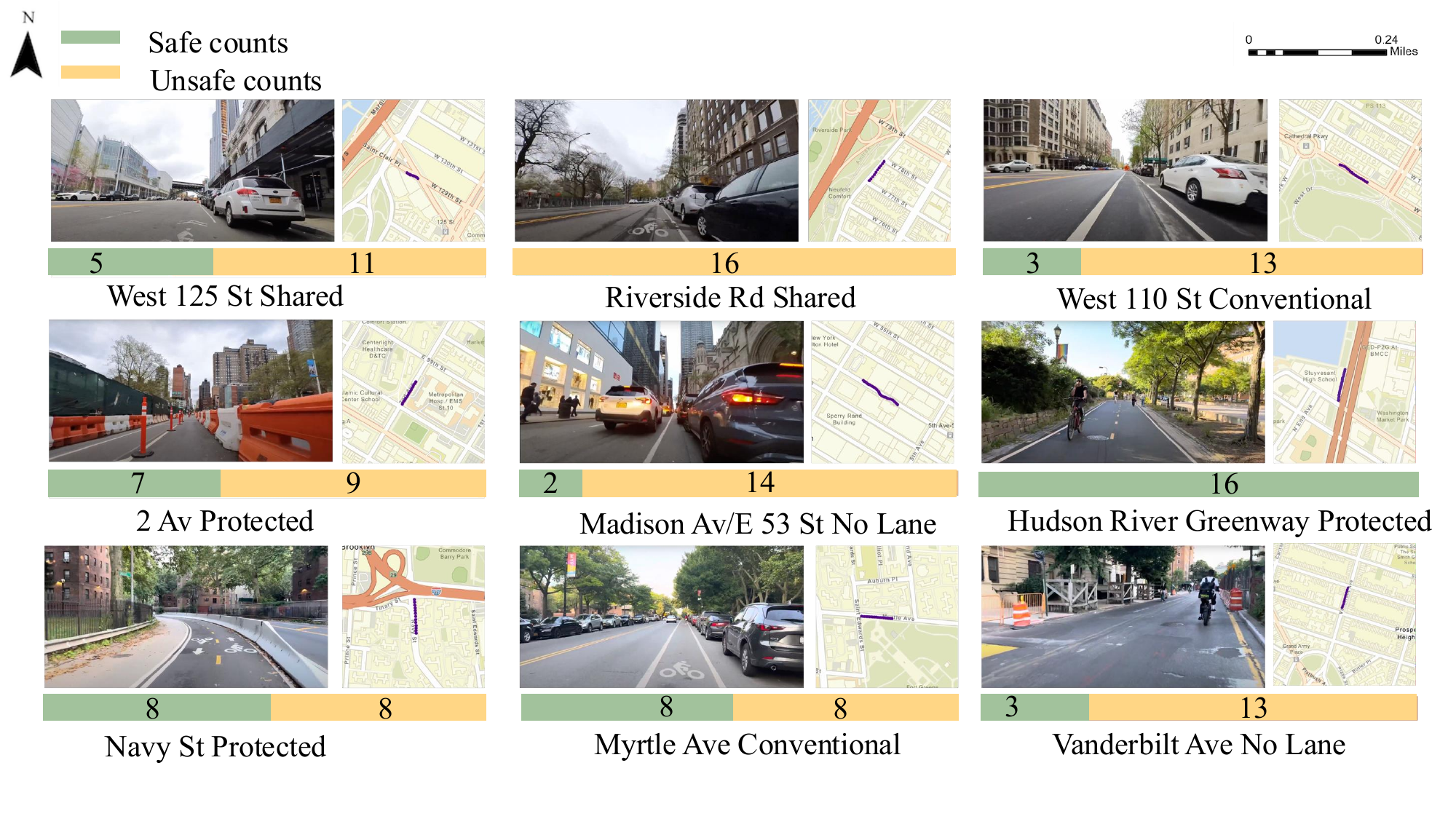} 
    \caption{Counts of "safe" and "unsafe" across the nine selected street segments. For each street segment, the left image shows a representative frame captured from the corresponding video, while the right image displays the mapped cycling trajectory of that same segment within the street network. Beneath each pair, green and yellow bars report the absolute counts of "safe" and "unsafe" responses, respectively.}
    \label{fig:9pic}
\end{figure}

For each of the nine selected street segments, Figure \ref{fig:9pic} displays the quantitative counts of "safe" and "unsafe" perceptions. Protected lanes, such as the Hudson River Greenway, received exclusively "safe" ratings, primarily attributed to advantageous bike lane width, surrounding greenery, and road surface quality. Conversely, roads without bike lanes, including Vanderbilt Ave and Madison Av/E 53 St, recorded the relatively highest "unsafe" counts (13 and 14), driven by road construction and the lack of dedicated cycling infrastructure. Conventional lanes yielded mixed results: on West 110 St, safety perceptions were undermined by side parking and construction despite decent road surface quality, while Myrtle Ave participants weighed concerns over narrow lanes and motor vehicles against benefits like greenery and low pedestrian volume. Notably, shared lanes such as Riverside Dr and West 125 St were perceived as predominantly "unsafe" (16 and 11 counts),  with Riverside Dr recording the highest number of “unsafe” responses among all segments, suggesting that the absence of dedicated cycling infrastructure and perceived exposure to motor traffic may outweigh otherwise favorable environmental conditions. Surprisingly, protected lanes also recorded relatively high “unsafe” counts in the presence of construction (e.g., 2 Av), where temporary disruptions coincide with reduced perceived safety despite the availability of dedicated cycling infrastructure.

\subsection{Feature Analysis}
\label{subsec:Feature_Analysis}

\begin{figure}[t!]
    \centering
    {%
    \includegraphics[width=1\textwidth]{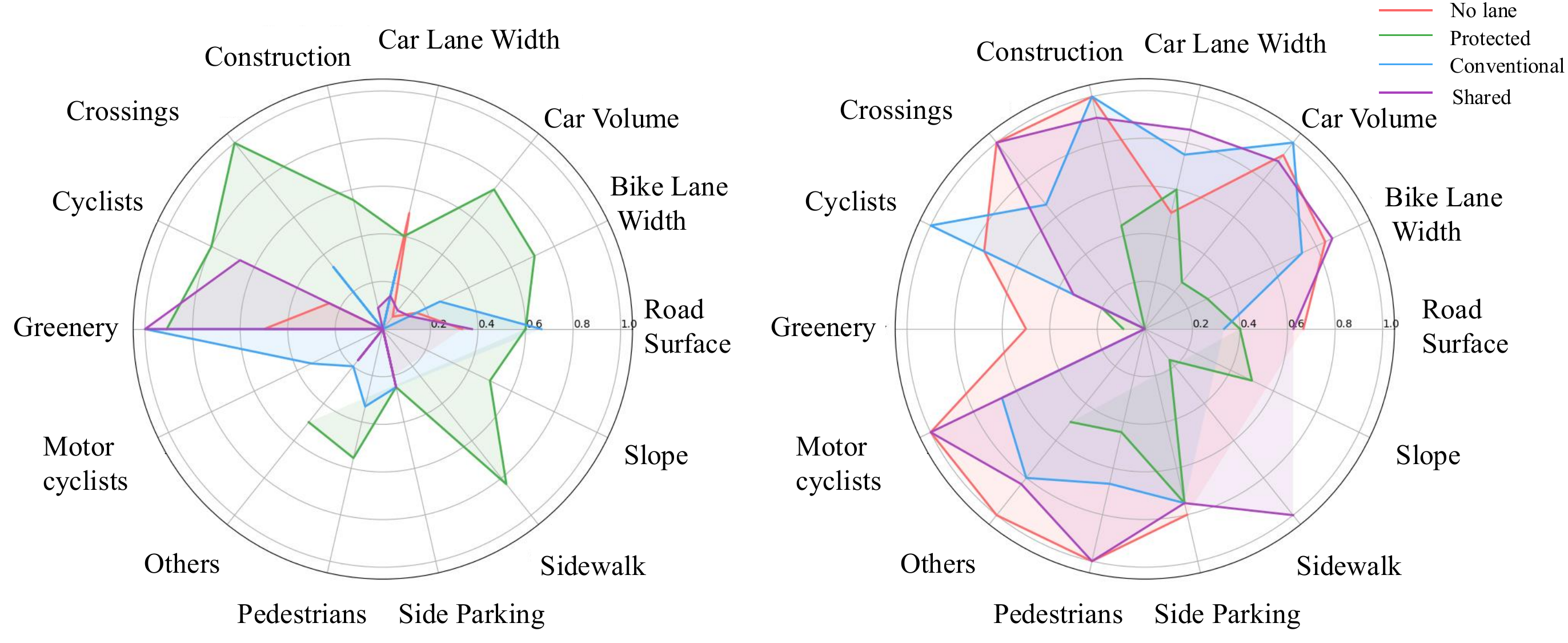}
    }
    \caption{Comparison of "safe" (left) and "unsafe" (right) feature ratios across four bike lane types (no lane, shared, conventional, and protected). Each radar axis represents a built environment or traffic feature, and the radius indicates the relative frequency with which participants referred to that feature when describing perceived safety or risk.}
    \label{fig:safe_unsafe_radar}
\end{figure}

Figure~\ref{fig:safe_unsafe_radar} compares the distribution of perceived "safe" and "unsafe" features across four types of bike lanes (no lane, shared, conventional, and protected). Positive safety cues are unevenly distributed and concentrate on a small number of features, with greenery emerging as the most consistently cited contributor to perceived safety across all lane types, particularly in protected and conventional settings. The presence of other cyclists is also commonly interpreted as reassuring rather than threatening, especially in protected and shared lanes, while infrastructure attributes such as bike lane width and road surface quality contribute to perceived safety mainly in conventional lanes. In contrast, unsafe perceptions are more widespread and dominated by traffic interactions. Across all lane types, motor vehicles, crossings, and motorcyclists constitute the primary sources of perceived risk, with particularly high concentrations on streets with no lanes or conventional lanes, reflecting greater exposure to mixed traffic and potential conflicts. Protected bike lanes consistently exhibit the lowest prevalence of unsafe features, underscoring their effectiveness in reducing perceived risk by limiting interactions with motor vehicles and other road users.

\subsection{Semantic Analysis}
\label{subsec:semantic}
The clustering analysis can be applied to the free text phrases extracted from the descriptions presented in Table \ref{tab:clusters_phrase_examples}. As a result, the clusters reveal higher level themes that are closely aligned with, but not identical to, the original 13 features. For example, one cluster groups phrases related to visibility and driveways (“tree blocking view”, “parking lot exiting”), bike lane presence and width (e.g., “separate bike lane”, “bike lane narrow”, “separation bike lane from car lanes”), while another captures side parking and mixed traffic (“cars parked”, “busy street”, “pedestrians jaywalking”). Other clusters correspond to road surface quality (“pavement smooth”, “cracks in the road”, “bumpy”), physical barriers, and construction (“construction blocking bike lane”, “set barrier”). This alignment is not accidental, it reflects that participants are reasoning in terms of the same physical dimensions that our feature coding scheme assumed, which strengthens the construct validity of our feature set. Beyond simple replication, these clusters differentiate distinct safety mechanisms within coded features, such as distinguishing  context-specific safety mechanisms within broader categories. By merging multiple attributes into emergent constructs like street legibility or exposure to traffic, this analysis reveals how riders articulate complex safety reasoning that transcends a simple counting of features. The findings in this section demonstrate that unstructured human–chatbot dialogues reflect consistent and interpretable safety themes, thereby fulfilling the third research objective of this study.

\begin{table}[t!]
\centering
\caption{Semantic clusters of perceived features and representative phrases}
\label{tab:clusters_phrase_examples}
\begin{adjustbox}{max width=\textwidth}
\begin{tabular}{clp{9cm}}
\toprule
\textbf{Cluster} & \textbf{Label} & \textbf{Example phrases (original wording)} \\
\midrule
0 & Driveways, cars and visibility 
  & Cars drive; parking lot exiting; driveway entrance; people cars drive; essentially two driveways; tree blocking view \\
\midrule
1 & Bike lane presence and width 
  & Bike lane; bike lanes; bike lane width makes feel; bike lane narrow; separation bike lane car lanes; narrow bike lane doesn \\
\midrule
2 & Traffic operations and transitions 
  & Need improve; suddenly start; crossroad stressful; illegal space; unstable factors; doesn continue \\
\midrule
3 & Side parking and mixed traffic 
  & Cars parked; busy street; let people walk; pedestrians jaywalking; lot parking; sidewalk wide clearly separated \\
\midrule
4 & Space, clutter and comfort 
  & Lane strip; people outdoors; amount trash; space narrow; median small; lot visual clutter \\
\midrule
5 & Dedicated cycling space and riders 
  & Separate bike; separate bike lane; bike share lane; crossing bikes crosswalk; car doesn affect bike; person riding bike lane \\
\midrule
6 & Physical barriers and protection 
  & Set barrier; guardrail added; physical barriers; barrier line; construction blocking bike lane; barrier protects cyclists \\
\midrule
7 & Markings, lights and signage 
  & Line clearly; pavement markings; traffic lights; bright lights; lane visually; painted bike lane \\
\midrule
8 & Surface quality and smoothness 
  & Pavement smooth; smooth road; cracks road; bumpy; potholes; uneven surface \\
\bottomrule
\end{tabular}
\end{adjustbox}
\begin{tablenotes}
    \footnotesize
    \item Clusters were derived from participant responses collected through the chatbot-based video evaluation. Text fragments were converted into semantic embeddings and grouped using an unsupervised clustering procedure. The number of clusters was determined based on thematic coherence. Cluster labels were assigned through qualitative interpretation of dominant semantic patterns. Example phrases are presented in their original wording without grammatical correction.
\end{tablenotes}
\end{table}
\FloatBarrier

\subsection{Mixed-effects Logistic Regression}
\label{subsec:regression}
Moreover, Table~\ref{tab:melr_features_ci_135} reports the results of a logistic regression in which the dependent variable indicates whether a given video section was classified as "safe". The model jointly includes basic socio-demographics, self-reported knowledge of cycling laws, and counts of how often each built environment feature was mentioned in the chat as a reason for feeling safe or unsafe. Model fit statistics are AIC = $163.4$, BIC = $241.8$, and log-likelihood = $-54.7$.

\begin{table}[ht]
\centering
\caption{Mixed-effects logistic regression for perceived safety (N=135)}
\label{tab:melr_features_ci_135}
\begin{adjustbox}{max width=\textwidth}
\begin{tabular}{lrrrrrr}
\toprule
& \multicolumn{3}{c}{Logit coefficient} & \multicolumn{3}{c}{Marginal effect} \\
\cmidrule(lr){2-4}\cmidrule(lr){5-7}
Variable & Coef. & 95\% CI & $p$-value & dy/dx & 95\% CI & $p$-value \\
\midrule
\multicolumn{7}{l}{\textbf{Perception scales and demographics}} \\
Cycling frequency: daily (vs.\ 2--6 times/week) & $1.72$ & [$-3.91,\ 7.35$] & $0.550$ & $0.22$ & [$-0.35,\ 0.79$] & $0.446$ \\
Cycling frequency: less than once/week (vs.\ 2--6 times/week) & $1.02$ & [$-2.46,\ 4.50$] & $0.567$ & $0.13$ & [$-0.22,\ 0.47$] & $0.479$ \\
Male (vs.\ female) & $0.88$ & [$-1.44,\ 3.21$] & $0.456$ & $0.12$ & [$-0.15,\ 0.39$] & $0.371$ \\
Age 25--34 (vs.\ 18--24) & $-0.16$ & [$-2.47,\ 2.14$] & $0.891$ & $-0.02$ & [$-0.28,\ 0.24$] & $0.875$ \\
Age 35--44 (vs.\ 18--24) & $3.68^{\dagger}$ & [$-0.12,\ 7.48$] & $0.058$ & $0.46^{**}$ & [$0.14,\ 0.79$] & $0.005$ \\
Bachelor's degree (vs.\ associates/technical degree) & $-1.49$ & [$-4.33,\ 1.36$] & $0.306$ & $-0.20$ & [$-0.53,\ 0.13$] & $0.232$ \\
Graduate/professional degree (vs.\ associates/technical degree) & $-0.30$ & [$-3.35,\ 2.75$] & $0.847$ & $-0.04$ & [$-0.41,\ 0.32$] & $0.820$ \\
Some high school or less (vs.\ associates/technical degree) & $0.43$ & [$-3.04,\ 3.89$] & $0.809$ & $0.06$ & [$-0.35,\ 0.47$] & $0.773$ \\
White or Caucasian (vs.\ Asian) & $-1.26$ & [$-3.13,\ 0.60$] & $0.185$ & $-0.16$ & [$-0.36,\ 0.04$] & $0.112$ \\
Knowledge of cycling laws & $-0.56$ & [$-1.54,\ 0.42$] & $0.264$ & $-0.08$ & [$-0.20,\ 0.04$] & $0.210$ \\
\midrule
\multicolumn{7}{l}{\textbf{Built-environment features}} \\
Road surface (O) & $0.24$ & [$-0.47,\ 0.96$] & $0.506$ & $0.03$ & [$-0.05,\ 0.12$] & $0.459$ \\
Bike lane width (O) & $-0.26$ & [$-0.92,\ 0.39$] & $0.432$ & $-0.04$ & [$-0.12,\ 0.05$] & $0.386$ \\
Car volume (O) & $-0.88^{*}$ & [$-1.72,\ -0.05$] & $0.039$ & $-0.12^{*}$ & [$-0.22,\ -0.02$] & $0.016$ \\
Car lane width (O) & $0.16$ & [$-0.70,\ 1.01$] & $0.720$ & $0.02$ & [$-0.09,\ 0.13$] & $0.693$ \\
Construction/barrier on bike lane (B) & $-1.35^{**}$ & [$-2.34,\ -0.36$] & $0.008$ & $-0.18^{**}$ & [$-0.30,\ -0.07$] & $0.001$ \\
Crossing (crossroads) (B) & $-0.62$ & [$-1.78,\ 0.54$] & $0.295$ & $-0.08$ & [$-0.22,\ 0.05$] & $0.234$ \\
Cyclists (B) & $0.77^{\dagger}$ & [$-0.09,\ 1.63$] & $0.078$ & $0.11^{*}$ & [$0.00,\ 0.21$] & $0.046$ \\
Greenery (O) & $1.42^{**}$ & [$0.35,\ 2.50$] & $0.010$ & $0.19^{**}$ & [$0.07,\ 0.32$] & $0.002$ \\
Motorcyclists (B) & $-1.18$ & [$-2.79,\ 0.42$] & $0.149$ & $-0.16$ & [$-0.36,\ 0.03$] & $0.107$ \\
Pedestrians (B) & $-0.93^{*}$ & [$-1.78,\ -0.07$] & $0.033$ & $-0.13^{*}$ & [$-0.23,\ -0.03$] & $0.014$ \\
Side parking (B) & $-0.04$ & [$-0.92,\ 0.84$] & $0.934$ & $-0.01$ & [$-0.11,\ 0.10$] & $0.926$ \\
Sidewalk (B) & $0.43$ & [$-0.96,\ 1.82$] & $0.542$ & $0.06$ & [$-0.11,\ 0.23$] & $0.492$ \\
Road grade (O) & $-1.00$ & [$-4.11,\ 2.11$] & $0.529$ & $-0.14$ & [$-0.52,\ 0.25$] & $0.490$ \\
Others & $-0.30$ & [$-2.06,\ 1.46$] & $0.736$ & $-0.04$ & [$-0.26,\ 0.18$] & $0.710$ \\
\midrule
\multicolumn{7}{l}{\textbf{Random effects}} \\
User ID variance (SD) & \multicolumn{6}{l}{$0e+00\ (0.0000)$} \\
Section variance (SD) & \multicolumn{6}{l}{$2.824\ (1.6803)$} \\
\midrule
Observations & \multicolumn{6}{c}{$135$} \\
Groups: User ID / Section & \multicolumn{6}{c}{$15$ / $9$} \\
Log-likelihood & \multicolumn{6}{c}{$-54.7$} \\
AIC & \multicolumn{6}{c}{$163.4$} \\
BIC & \multicolumn{6}{c}{$241.8$} \\
\bottomrule
\end{tabular}
\end{adjustbox}
\begin{flushleft}
\footnotesize
Notes: Dependent variable equals 1 if a section was classified as ``safe''. Coefficients are from a mixed-effects logistic regression with random intercepts for User ID and section. dy/dx reports average marginal effects from fixed-effects-only predicted probabilities. Coefficient 95\% CIs are approximated as estimate $\pm 1.96\times$SE. One fixed-effect column (Q9 Others/Unknown) was dropped due to rank deficiency. Variable types are indicated as (B) for binary and (O) for ordinal-scale measures, consistent with the notation introduced in Table \ref{tab:feature_eva}.
$^{\dagger} p<0.10,\ ^{*} p<0.05,\ ^{**} p<0.01,\ ^{***} p<0.001$.
\end{flushleft}
\label{tab:melr_features_ci_135}
\end{table}

The results from the mixed-effects logistic regression support the feasibility of the AI chatbot workflow for collecting analyzable urban perception data. Among the fixed effects, greenery is positively associated with perceived safety and has larger marginal effects than the built-environment variables. The presence of greenery is associated with an approximately $0.19$ higher predicted probability. Conversely, car volume and pedestrian presence are associated with a $0.12$ and $0.13$ lower predicted probability of classifying a section as safe. The presence of cyclists is also positively associated with safety judgments, whereas construction-related obstructions show a negative association. Other traffic-related features, such as street parking and motorcyclists, also show negative associations. In our sampling, demographic contrasts suggest that age may be associated with perceived safety, although other demographic effects are not statistically significant. Given the pilot nature of the study, these findings should be interpreted as preliminary evidence of analytical feasibility rather than behavioral conclusions. The consistency of these results with established urban design theories supports our developed video-based human–chatbot method.

\FloatBarrier

\subsection{Chatbot Feedback}
\begin{figure}[ht]
    \centering
    {%
    \includegraphics[width=\textwidth]{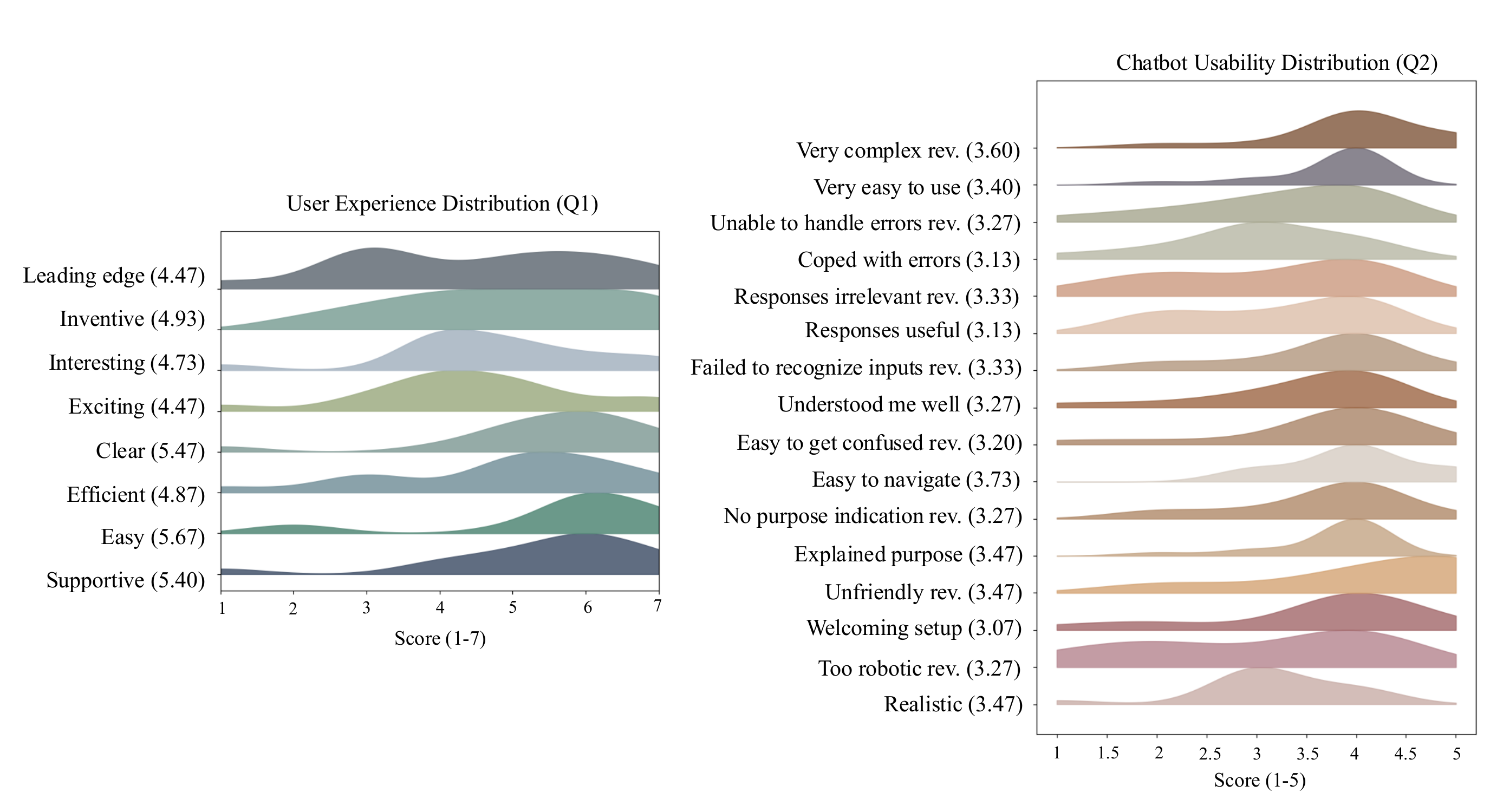}
    }
    \caption{The left panel presents the distribution of semantic differential ratings measuring User Experience (Q1), while the right panel shows the distribution of Likert-scale responses assessing Chatbot Usability (Q2). Each horizontal density curve represents one evaluation item. Each horizontal curve represents a kernel density estimate (KDE) of the observed discrete rating data. Values in parentheses indicate the mean score for each item; items marked as “rev.” were reverse-coded prior to analysis.}
    \label{fig:chatbotevaluationscores}
\end{figure}

Feedback on chatbot use was also collected in Figure \ref{fig:chatbotevaluationscores}. Participants evaluated the system using a seven-item semantic differential scale, and the results indicate an overall positive reception. On average, respondents rated the AI chatbot with mean scores of 5.00 out of 7 (standard deviation = 1.60), reflecting generally favorable evaluations across dimensionts such as Easy, Clear, Supportive. In terms of chatbot usability, to ensure a coherent analysis, items with inverted scale directions or negative phrasing were reverse-coded so that higher scores consistently indicated more positive evaluations. The overall mean score was 3.47 out of 5 (standard deviation = 0.43). Relatively higher scores were observed for items such as ease of navigation, clarity of purpose (explained purpose), and perceived realism. Notably, negative phrased factors such as perceived roboticness, input recognition failure, or excessive complexity received moderate scores after reverse coding, indicating that respondents did not strongly perceive these as major issues. Collectively, these findings demonstrate that the integrated video-based AI chatbot survey framework is a feasible, reliable, and engaging tool for capturing high-quality situational perceptions, thereby successfully fulfilling the first research objective of this study.

\FloatBarrier


\section{Discussion}
This study embeds a generative AI chatbot into an online survey to elicit individuals’ perspectives on cycling safety in New York City. The chatbot was designed to capture both overall safety judgments and feature-specific perceptions through interactive dialog, while the survey provided an external benchmark for evaluating the chatbot’s measurement performance. This mixed approach enables triangulation between conventional instrument-based measurement and conversational elicitation, while offering a scalable pathway for collecting qualitative safety perceptions that are challenging to obtain through fixed-response items alone.

The first objective was to assess the feasibility of the chatbot-based survey framework. This feasibility was explicitly measured through the survey component, including user feedback and interaction level responses collected during the chatbot-mediated tasks. The successful deployment demonstrates that a video-based AI interface can technically function as a robust data collection instrument. However, the study also revealed significant insights regarding participant engagement and survey design. Initially, 41 responses were recorded, however, after excluding internal test samples and incomplete submissions, only 16 unique participants (39.0\%) remained who completed the full evaluation of all nine video segments. This high attrition rate highlights a critical trade-off between data depth and user retention. The current experimental design required participants to engage in structured dialogue for nine consecutive videos, imposing a substantial cognitive load and leading to survey fatigue. While this repetitive process limited the final sample size, the data obtained from the 16 completers is exceptionally dense and of high-quality, validating the internal feasibility of the method. The framework proved capable of eliciting detailed reasoning, however, future deployments must optimize for engagement, likely by adopting a "distributed sampling" strategy \citep{raghunathanSplitQuestionnaireSurvey1995}, where participants evaluate a smaller subset of videos (e.g., 1–2 segments) to maintain interest and reduce drop-off rates.

Regarding the second objective,  the feasibility of the video-integrated chatbot-based survey was evaluated through descriptive statistics of user feedback. The overall mean scores indicate a generally positive reception, suggesting that the AI chatbot was perceived as usable and moderately engaging. However, the distribution of responses reveals specific refinements. In user experience, the density curves for "Interesting" and "Exciting" are much flatter and spread toward the middle. Similarly, the feedback on chatbot usability shows a noticeable dip in "Welcoming setup" and a cluster of scores flagging the bot as "Too robotic". These visual trends in the data point to a clear mandate for the next iteration: the technical foundation is solid, but the conversational layer needs more warmth and diversified phrasing to keep users from feeling like they are just interacting with a robot.

The findings from the mixed-effects logistic regression show that the potential of human-chatbot conversations can identify the attributes that drive safety perceptions. As a pilot study, preliminary results from data collected from 16 participants serve as proof of concept for the methodology. Specifically, the model reveals that greenery and the presence of other cyclists show positive correlations, while construction obstacles and motor vehicle traffic show negative correlations. Roadside parking and motorcycle riders exhibit additional negative effects approaching conventional significance thresholds. This supports the argument that the chatbot can reveal substantive feature-level mechanisms underlying perceived cycling safety, rather than merely replicating generalized attitudes.


The semantic analysis confirms that unstructured human–chatbot dialogues reflect consistent and interpretable safety reasoning. The clustering results demonstrate that participants' free-text responses are not random noise but closely map to physical street features while adding crucial contextual nuance. For instance, the system successfully distinguished between different types of risks associated with "side parking". Moreover, suggestions such as “bright lights” or “set barrier” emerged organically in the dialogues, indicating that chatbot-embedded surveys can surface bottom-up safety concerns in a form that is interpretable by planners and decision-makers, while simultaneously enabling individuals to directly communicate context-specific feedback to those responsible for cycling safety governance.

Despite these contributions, several limitations should be acknowledged, which also indicate directions for future research. As mentioned, the sample size could be improved by employing distributed or adaptive sampling strategies. Second, heterogeneity in perceived safety may partly reflect unmeasured individual and context-level variations, such as weather \citep{NANKERVIS1999417} and seasonality \citep{KUMMENEJE201940}，which can systematically influence cyclists’ risk perception but were not explicitly captured in the present design.

Furthermore, as a pilot study, this work is primarily intended to demonstrate methodological feasibility rather than exhaustive spatial coverage. Future work could build on this framework by integrating objective video sensing from existing infrastructure, such as CCTV, to enable large-scale deployment and improve external validity. As demonstrated by \cite{Zangen2016}, video-based frameworks can objectively quantify cyclist safety risks through dynamic metrics like kinetic energy. In addition, large-scale street-level imagery (e.g., via the Google Street View API) and computer-vision–based extraction of built-environment attributes could further support systematic and reproducible objective evaluation across diverse urban settings \citep{BILJECKI2021104217}. Correlating these objective traffic observations with the subjective perceptions captured by the chatbot would establish a more robust, dual-validated model of urban cycling safety that is scalable beyond the specific context.

\section{Conclusion}

Overall, this pilot study highlights the promise of AI-driven methodologies for urban mobility research by demonstrating how an interactive video-based AI chatbot can capture both general cycling safety perceptions and feature-level explanations. The findings indicate that all three research objectives have been achieved, including the development of a controllable conversational workflow, the evaluation of both method and data feasibility, and the assessment of using human–chatbot textual data to capture perceived cycling safety. While preliminary, these findings suggest that the proposed workflow may provide a replicable approach for studying perceptions of cycling safety across diverse urban contexts.

\section*{Funding}

This research did not receive any external funding.

\section*{Data Ethics}
This study was approved by the Institutional Review Board (IRB) of New York University under protocol number IRB-FY2025-10241. All participants were informed of the study's purpose and their right to withdraw at any time.

\section*{Data Availability Statement}

The participants in this study did not provide written consent for their data to be shared publicly; therefore, due to the sensitive nature of the research, supporting data is not available.

\section*{Acknowledgments}

The authors sincerely thank the following students from New York University for their assistance with data collection: PhD candidate Yin Wen, and Master’s students Lishun Liu and Sizhe Xu. We also thank all study participants for their valuable support and involvement.

\section*{Use of Generative AI}

The authors used a generative artificial intelligence (AI) tool (ChatGPT, OpenAI) to assist with language editing, minor formatting improvements, and limited coding support (e.g., drafting and troubleshooting code) during manuscript preparation. The AI system was not used for data generation, analysis, interpretation, or decision-making. All analytical decisions, code, and final content were reviewed and approved by the authors, who take full responsibility for the integrity of the work.

\clearpage
\bibliographystyle{apalike-refs}
\bibliography{bib}

@misc{nycDOTCycling,
  author = {{New York City Department of Transportation}},
  title = {Cycling in the City},
  howpublished = {\url{https://www.nyc.gov/html/dot/html/bicyclists/cyclinginthecity.shtml}},
  year = {2024}, 
  note = {Accessed: 2024-03-09}
}

@article{Skoczynski2021,
  author = {Skoczyński, P.},
  title = {Analysis of Solutions Improving Safety of Cyclists in the Road Traffic},
  journal = {Applied Sciences},
  volume = {11},
  number = {9},
  pages = {3771},
  year = {2021},
  doi = {10.3390/app11093771},
  url = {https://doi.org/10.3390/app11093771},

}

@misc{NYCDCPDigitalMap,
  author = {{New York City Department of City Planning}},
  title = {Digital City Map (DCM)},
  year = {2024}, 
  url = {https://www.nyc.gov/site/planning/data-maps/open-data/dwn-digital-city-map.page},
}

@misc{NYCStreetDesignBikeLane2020,
  author = {{New York City Department of Transportation}},
  title = {Bike Lane Design Guidelines},
  year = {2020},
  url = {https://www.nycstreetdesign.info/sites/default/files/2020-03/Bike-Lane-table.pdf},
}

@Article{safety9040075,
AUTHOR = {Duren, Michelle and Corrigan, Bryce and Kennedy, Ryan David and Pollack Porter, Keshia M. and Ehsani, Johnathon},
TITLE = {Identifying and Assessing Perceived Cycling Safety Components},
JOURNAL = {Safety},
VOLUME = {9},
YEAR = {2023},
NUMBER = {4},
ARTICLE-NUMBER = {75},
URL = {https://www.mdpi.com/2313-576X/9/4/75},
ISSN = {2313-576X},
DOI = {10.3390/safety9040075}
}

@article{JEON2024103992,
  title = {The effects of built environments on bicycle accidents around bike-sharing program stations using street view images and deep learning techniques: The moderating role of streetscape features},
  volume = {121},
  ISSN = {0966-6923},
  url = {http://dx.doi.org/10.1016/j.jtrangeo.2024.103992},
  DOI = {10.1016/j.jtrangeo.2024.103992},
  journal = {Journal of Transport Geography},
  publisher = {Elsevier BV},
  author = {Jeon,  Junehyung and Woo,  Ayoung},
  year = {2024},
  month = dec,
  pages = {103992}
}

@article{BLITZ202127,
  title = {How does the individual perception of local conditions affect cycling? An analysis of the impact of built and non-built environment factors on cycling behaviour and attitudes in an urban setting},
  volume = {25},
  ISSN = {2214-367X},
  url = {http://dx.doi.org/10.1016/j.tbs.2021.05.006},
  DOI = {10.1016/j.tbs.2021.05.006},
  journal = {Travel Behaviour and Society},
  publisher = {Elsevier BV},
  author = {Blitz,  Andreas},
  year = {2021},
  month = oct,
  pages = {27–40}
}

@article{AIApproach2019,
  title = {An AI Approach to Collecting and Analyzing Human Interactions With Urban Environments},
  volume = {7},
  ISSN = {2169-3536},
  url = {http://dx.doi.org/10.1109/ACCESS.2019.2943845},
  DOI = {10.1109/access.2019.2943845},
  journal = {IEEE Access},
  publisher = {Institute of Electrical and Electronics Engineers (IEEE)},
  author = {Ferrara,  Enrico and Fragale,  Luigi and Fortino,  Giancarlo and Song,  Wei and Perra,  Cristian and Di Mauro,  Mario and Liotta,  Antonio},
  year = {2019},
  pages = {141476–141486}
}

@article{Senadheera01022024,
  title = {Understanding Chatbot Adoption in Local Governments: A Review and Framework},
  volume = {32},
  ISSN = {1466-1853},
  url = {http://dx.doi.org/10.1080/10630732.2023.2297665},
  DOI = {10.1080/10630732.2023.2297665},
  number = {3},
  journal = {Journal of Urban Technology},
  publisher = {Informa UK Limited},
  author = {Senadheera,  Sajani and Yigitcanlar,  Tan and Desouza,  Kevin C. and Mossberger,  Karen and Corchado,  Juan and Mehmood,  Rashid and Li,  Rita Yi Man and Cheong,  Pauline Hope},
  year = {2024},
  month = feb,
  pages = {35–69}
}

@article{Ma2014,
  title = {The objective versus the perceived environment: what matters for bicycling?},
  volume = {41},
  ISSN = {1572-9435},
  url = {http://dx.doi.org/10.1007/s11116-014-9520-y},
  DOI = {10.1007/s11116-014-9520-y},
  number = {6},
  journal = {Transportation},
  publisher = {Springer Science and Business Media LLC},
  author = {Ma,  Liang and Dill,  Jennifer and Mohr,  Cynthia},
  year = {2014},
  month = apr,
  pages = {1135–1152}
}

@article{ANDROUTSOPOULOU2019358,
  title = {Transforming the communication between citizens and government through AI-guided chatbots},
  volume = {36},
  ISSN = {0740-624X},
  url = {http://dx.doi.org/10.1016/j.giq.2018.10.001},
  DOI = {10.1016/j.giq.2018.10.001},
  number = {2},
  journal = {Government Information Quarterly},
  publisher = {Elsevier BV},
  author = {Androutsopoulou,  Aggeliki and Karacapilidis,  Nikos and Loukis,  Euripidis and Charalabidis,  Yannis},
  year = {2019},
  month = apr,
  pages = {358–367}
}

@article{Henman01102020,
  title = {Improving public services using artificial intelligence: possibilities,  pitfalls,  governance},
  volume = {42},
  ISSN = {2327-6673},
  url = {http://dx.doi.org/10.1080/23276665.2020.1816188},
  DOI = {10.1080/23276665.2020.1816188},
  number = {4},
  journal = {Asia Pacific Journal of Public Administration},
  publisher = {Informa UK Limited},
  author = {Henman,  Paul},
  year = {2020},
  month = sep,
  pages = {209–221}
}

@misc{grootendorst2020keybert,
  author       = {Maarten Grootendorst},
  title        = {KeyBERT: Minimal keyword extraction with BERT.},
  year         = 2020,
  publisher    = {Zenodo},
  version      = {v0.3.0},
  doi          = {10.5281/zenodo.4461265},
  url          = {https://doi.org/10.5281/zenodo.4461265}
}

@article{ETMINANIGHASRODASHTI2018241,
  title = {Recreational cycling in a coastal city: Investigating lifestyle,  attitudes and built environment in cycling behavior},
  volume = {39},
  ISSN = {2210-6707},
  url = {http://dx.doi.org/10.1016/j.scs.2018.02.037},
  DOI = {10.1016/j.scs.2018.02.037},
  journal = {Sustainable Cities and Society},
  publisher = {Elsevier BV},
  author = {Etminani-Ghasrodashti,  Roya and Paydar,  Mohammad and Ardeshiri,  Ali},
  year = {2018},
  month = may,
  pages = {241–251}
}

@article{DAI2021101013,
  title = {Review of contextual elements affecting bicyclist safety},
  volume = {20},
  ISSN = {2214-1405},
  url = {http://dx.doi.org/10.1016/j.jth.2021.101013},
  DOI = {10.1016/j.jth.2021.101013},
  journal = {Journal of Transport \& Health},
  publisher = {Elsevier BV},
  author = {Dai,  Boya and Dadashova,  Bahar},
  year = {2021},
  month = mar,
  pages = {101013}
}

@article{Hasan-2023,
  title = {Specialized Urban Planning Chatbot: A Participatory Approach for Evaluating Efficiency},
  volume = {10},
  ISSN = {2012-5720},
  url = {http://dx.doi.org/10.4038/bhumi.v10i2.106},
  DOI = {10.4038/bhumi.v10i2.106},
  number = {2},
  journal = {Bhumi,  The Planning Research Journal},
  publisher = {Sri Lanka Journals Online},
  author = {Hasan,  Farasath and Jayasinghe,  Amila and Sathsarana,  Sahan},
  year = {2023},
  month = dec,
  pages = {3–16}
}

@article{Peng2024,
author = {Zhong-Ren Peng and Kai-Fa Lu and Yanghe Liu and Wei Zhai},
title ={The Pathway of Urban Planning AI: From Planning Support to Plan-Making},

journal = {Journal of Planning Education and Research},
volume = {44},
number = {4},
pages = {2263-2279},
year = {2024},
doi = {10.1177/0739456X231180568},

URL = { 
    
        https://doi.org/10.1177/0739456X231180568
    
    

},
eprint = { 
    
        https://doi.org/10.1177/0739456X231180568
    
    

}
}

@article{HERATH2022100076,
  title = {Adoption of artificial intelligence in smart cities: A comprehensive review},
  volume = {2},
  ISSN = {2667-0968},
  url = {http://dx.doi.org/10.1016/j.jjimei.2022.100076},
  DOI = {10.1016/j.jjimei.2022.100076},
  number = {1},
  journal = {International Journal of Information Management Data Insights},
  publisher = {Elsevier BV},
  author = {Herath,  H.M.K.K.M.B. and Mittal,  Mamta},
  year = {2022},
  month = apr,
  pages = {100076}
}

@article{BI2023103551,
  title = {Bicycle safety outside the crosswalks: Investigating cyclists’ risky street-crossing behavior and its relationship with built environment},
  volume = {108},
  ISSN = {0966-6923},
  url = {http://dx.doi.org/10.1016/j.jtrangeo.2023.103551},
  DOI = {10.1016/j.jtrangeo.2023.103551},
  journal = {Journal of Transport Geography},
  publisher = {Elsevier BV},
  author = {Bi,  Hui and Li,  Aoyong and Zhu,  He and Ye,  Zhirui},
  year = {2023},
  month = apr,
  pages = {103551}
}

@article{KWON2020105716,
  title = {An examination of the intersection environment associated with perceived crash risk among school-aged children: using street-level imagery and computer vision},
  volume = {146},
  ISSN = {0001-4575},
  url = {http://dx.doi.org/10.1016/j.aap.2020.105716},
  DOI = {10.1016/j.aap.2020.105716},
  journal = {Accident Analysis \& Prevention},
  publisher = {Elsevier BV},
  author = {Kwon,  Jae-Hong and Cho,  Gi-Hyoug},
  year = {2020},
  month = oct,
  pages = {105716}
}

@article{BI202251,
  title = {Examining the varying influences of built environment on bike-sharing commuting: Empirical evidence from Shanghai},
  volume = {129},
  ISSN = {0967-070X},
  url = {http://dx.doi.org/10.1016/j.tranpol.2022.10.004},
  DOI = {10.1016/j.tranpol.2022.10.004},
  journal = {Transport Policy},
  publisher = {Elsevier BV},
  author = {Bi,  Hui and Li,  Aoyong and Hua,  Mingzhuang and Zhu,  He and Ye,  Zhirui},
  year = {2022},
  month = dec,
  pages = {51–65}
}

@article{LAWSON2013499,
  title = {Perception of safety of cyclists in Dublin City},
  volume = {50},
  ISSN = {0001-4575},
  url = {http://dx.doi.org/10.1016/j.aap.2012.05.029},
  DOI = {10.1016/j.aap.2012.05.029},
  journal = {Accident Analysis \& Prevention},
  publisher = {Elsevier BV},
  author = {Lawson,  Anneka R. and Pakrashi,  Vikram and Ghosh,  Bidisha and Szeto,  W.Y.},
  year = {2013},
  month = jan,
  pages = {499–511}
}

@techreport{hunter2000evaluation,
  title = {Evaluation of a Combined Bicycle Lane/Right-Turn Lane in Eugene, Oregon},
  author = {Hunter, W. W.},
  editor = {{University of North Carolina (System). Highway Safety Research Center}},
  year = 2000,
  month = aug,
  edition = {Final Report October 1997 - July 1999},
  publisher = {{United States. Federal Highway Administration. Office of Safety Research and Development}},
  issn = {FHWA-RD-00-151},
  url          = {https://rosap.ntl.bts.gov/view/dot/40606}
}

@article{MINIKEL2012241,
  title = {Cyclist safety on bicycle boulevards and parallel arterial routes in Berkeley,  California},
  volume = {45},
  ISSN = {0001-4575},
  url = {http://dx.doi.org/10.1016/j.aap.2011.07.009},
  DOI = {10.1016/j.aap.2011.07.009},
  journal = {Accident Analysis \& Prevention},
  publisher = {Elsevier BV},
  author = {Minikel,  Eric},
  year = {2012},
  month = mar,
  pages = {241–247}
}

@article{tian2025quantifying,
  title = {Quantifying the non-isomorphism of global urban road networks using GNNs and graph kernels},
  volume = {15},
  ISSN = {2045-2322},
  url = {http://dx.doi.org/10.1038/s41598-025-90839-x},
  DOI = {10.1038/s41598-025-90839-x},
  number = {1},
  journal = {Scientific Reports},
  publisher = {Springer Science and Business Media LLC},
  author = {Tian,  Linfang and Rao,  Weixiong and Zhao,  Kai and Vo,  Huy T.},
  year = {2025},
  month = feb 
}

@article{Lusk2013,
  title = {Bicycle Guidelines and Crash Rates on Cycle Tracks in the United States},
  volume = {103},
  ISSN = {1541-0048},
  url = {http://dx.doi.org/10.2105/AJPH.2012.301043},
  DOI = {10.2105/ajph.2012.301043},
  number = {7},
  journal = {American Journal of Public Health},
  publisher = {American Public Health Association},
  author = {Lusk,  Anne C. and Morency,  Patrick and Miranda-Moreno,  Luis F. and Willett,  Walter C. and Dennerlein,  Jack T.},
  year = {2013},
  month = jul,
  pages = {1240–1248}
}

@techreport{huang1999comparative,
  title        = {A Comparative Analysis of Bicycle Lanes Versus Wide Curb Lanes},
  author       = {Huang, Herman H. and Hunter, Walter W. and Pein, Wayne E. and Stewart, J. Richard and Stutts, Jane C.},
  year         = {1999},
  month        = oct,
  institution  = {Federal Highway Administration},
  type         = {Technical Report},
  number       = {FHWA-RD-99-034},
  address      = {Washington, DC},
  note         = {Final Report},
  url          = {https://www.fhwa.dot.gov/publications/research/safety/pedbike/99034/99034.pdf}
}

@article{PUCHER2010S106,
title = {Infrastructure, programs, and policies to increase bicycling: An international review},
journal = {Preventive Medicine},
volume = {50},
pages = {S106-S125},
year = {2010},
issn = {0091-7435},
doi = {https://doi.org/10.1016/j.ypmed.2009.07.028},
url = {https://www.sciencedirect.com/science/article/pii/S0091743509004344},
author = {John Pucher and Jennifer Dill and Susan Handy}
}

@article{ZENG2024103739,
  title = {Measuring cyclists’ subjective perceptions of the street riding environment using K-means SMOTE-RF model and street view imagery},
  volume = {128},
  ISSN = {1569-8432},
  url = {http://dx.doi.org/10.1016/j.jag.2024.103739},
  DOI = {10.1016/j.jag.2024.103739},
  journal = {International Journal of Applied Earth Observation and Geoinformation},
  publisher = {Elsevier BV},
  author = {Zeng,  Qisheng and Gong,  Zheng and Wu,  Songtai and Zhuang,  Caigang and Li,  Shaoying},
  year = {2024},
  month = apr,
  pages = {103739}
}

@article{Yu2024StreetGreenery,
  author    = {Yu, Jiahua and Zhang, Hao and Dong, Xinyang and Shen, Jing},
  title     = {The impact of street greenery on active travel: a narrative systematic review},
  journal   = {Frontiers in Public Health},
  year      = {2024},
  volume    = {12},
  pages     = {1337804},
  doi       = {10.3389/fpubh.2024.1337804},
  pmid      = {38481839},
  pmcid     = {PMC10936756},
  url       = {https://doi.org/10.3389/fpubh.2024.1337804}
}

@article{Brownson2009,
  title = {Measuring the Built Environment for Physical Activity},
  volume = {36},
  ISSN = {0749-3797},
  url = {http://dx.doi.org/10.1016/j.amepre.2009.01.005},
  DOI = {10.1016/j.amepre.2009.01.005},
  number = {4},
  journal = {American Journal of Preventive Medicine},
  publisher = {Elsevier BV},
  author = {Brownson,  Ross C. and Hoehner,  Christine M. and Day,  Kristen and Forsyth,  Ann and Sallis,  James F.},
  year = {2009},
  month = apr,
  pages = {S99--S123.e12}
}

@article{Teschke2012slope,
  title = {Route Infrastructure and the Risk of Injuries to Bicyclists: A Case-Crossover Study},
  volume = {102},
  ISSN = {1541-0048},
  url = {http://dx.doi.org/10.2105/AJPH.2012.300762},
  DOI = {10.2105/ajph.2012.300762},
  number = {12},
  journal = {American Journal of Public Health},
  publisher = {American Public Health Association},
  author = {Teschke,  Kay and Harris,  M. Anne and Reynolds,  Conor C. O. and Winters,  Meghan and Babul,  Shelina and Chipman,  Mary and Cusimano,  Michael D. and Brubacher,  Jeff R. and Hunte,  Garth and Friedman,  Steven M. and Monro,  Melody and Shen,  Hui and Vernich,  Lee and Cripton,  Peter A.},
  year = {2012},
  month = dec,
  pages = {2336–2343}
}

@article{SCHIMEK2018sideparking,
  title = {Bike lanes next to on-street parallel parking},
  volume = {120},
  ISSN = {0001-4575},
  url = {http://dx.doi.org/10.1016/j.aap.2018.08.002},
  DOI = {10.1016/j.aap.2018.08.002},
  journal = {Accident Analysis \& Prevention},
  publisher = {Elsevier BV},
  author = {Schimek,  Paul},
  year = {2018},
  month = nov,
  pages = {74–82}
}

@article{Dozza2014cyclistpedestrain,
  title = {Understanding Bicycle Dynamics and Cyclist Behavior From Naturalistic Field Data (November 2012)},
  volume = {15},
  ISSN = {1558-0016},
  url = {http://dx.doi.org/10.1109/TITS.2013.2279687},
  DOI = {10.1109/tits.2013.2279687},
  number = {1},
  journal = {IEEE Transactions on Intelligent Transportation Systems},
  publisher = {Institute of Electrical and Electronics Engineers (IEEE)},
  author = {Dozza,  Marco and Fernandez,  Andre},
  year = {2014},
  month = feb,
  pages = {376–384}
}

@Article{Ivan2023motor,
AUTHOR = {Ivanišević, Tijana and Trifunović, Aleksandar and Čičević, Svetlana and Pešić, Dalibor and Simović, Sreten and Zunjic, Aleksandar and Duplakova, Darina and Duplak, Jan and Manojlovic, Uros},
TITLE = {Analysis and Determination of the Lateral Distance Parameters of Vehicles When Overtaking an Electric Bicycle from the Point of View of Road Safety},
JOURNAL = {Applied Sciences},
VOLUME = {13},
YEAR = {2023},
NUMBER = {3},
ARTICLE-NUMBER = {1621},
URL = {https://www.mdpi.com/2076-3417/13/3/1621},
ISSN = {2076-3417},
DOI = {10.3390/app13031621}
}

@article{MARSHALL2019types,
  title = {Why cities with high bicycling rates are safer for all road users},
  volume = {13},
  ISSN = {2214-1405},
  url = {http://dx.doi.org/10.1016/j.jth.2019.03.004},
  DOI = {10.1016/j.jth.2019.03.004},
  journal = {Journal of Transport \& Health},
  publisher = {Elsevier BV},
  author = {Marshall,  Wesley E. and Ferenchak,  Nicholas N.},
  year = {2019},
  month = jun,
  pages = {100539}
}

@article{nieuwenhuijsen2020urban,
  title = {Urban and transport planning pathways to carbon neutral,  liveable and healthy cities; A review of the current evidence},
  volume = {140},
  ISSN = {0160-4120},
  url = {http://dx.doi.org/10.1016/j.envint.2020.105661},
  DOI = {10.1016/j.envint.2020.105661},
  journal = {Environment International},
  publisher = {Elsevier BV},
  author = {Nieuwenhuijsen,  Mark J.},
  year = {2020},
  month = jul,
  pages = {105661}
}

@article{Ricchetti2025Greenery,
    author = {Ricchetti, Chiara and Rotaris, Lucia},
    title = {The role of linear green infrastructure in perceived cycling safety: insights from Berlin},
    journal = {Urbanization, Sustainability and Society},
    volume = {2},
    number = {1},
    pages = {338-363},
    year = {2025},
    month = {08},
    issn = {2976-8993},
    doi = {10.1108/USS-02-2025-0008},
    url = {https://doi.org/10.1108/USS-02-2025-0008},
    eprint = {https://www.emerald.com/uss/article-pdf/2/1/338/10339735/uss-02-2025-0008en.pdf},
}

@article{Jacobsen2003SafetyInNumbers,
  title = {Safety in numbers: more walkers and bicyclists,  safer walking and bicycling},
  volume = {9},
  ISSN = {1475-5785},
  url = {http://dx.doi.org/10.1136/ip.9.3.205},
  DOI = {10.1136/ip.9.3.205},
  number = {3},
  journal = {Injury Prevention},
  publisher = {BMJ},
  author = {Jacobsen,  P L},
  year = {2003},
  month = sep,
  pages = {205–209}
}

@article{KUMMENEJE201940,
  title = {Seasonal variation in risk perception and travel behaviour among cyclists in a Norwegian urban area},
  volume = {124},
  ISSN = {0001-4575},
  url = {http://dx.doi.org/10.1016/j.aap.2018.12.021},
  DOI = {10.1016/j.aap.2018.12.021},
  journal = {Accident Analysis \& Prevention},
  publisher = {Elsevier BV},
  author = {Kummeneje,  An-Magritt and Ryeng,  Eirin Olaussen and Rundmo,  Torbjørn},
  year = {2019},
  month = mar,
  pages = {40–49}
}

@article{NANKERVIS1999417,
  title = {The effect of weather and climate on bicycle commuting},
  volume = {33},
  ISSN = {0965-8564},
  url = {http://dx.doi.org/10.1016/S0965-8564(98)00022-6},
  DOI = {10.1016/s0965-8564(98)00022-6},
  number = {6},
  journal = {Transportation Research Part A: Policy and Practice},
  publisher = {Elsevier BV},
  author = {Nankervis,  Max},
  year = {1999},
  month = aug,
  pages = {417–431}
}

@article{BILJECKI2021104217,
  title = {Street view imagery in urban analytics and GIS: A review},
  volume = {215},
  ISSN = {0169-2046},
  url = {http://dx.doi.org/10.1016/j.landurbplan.2021.104217},
  DOI = {10.1016/j.landurbplan.2021.104217},
  journal = {Landscape and Urban Planning},
  publisher = {Elsevier BV},
  author = {Biljecki,  Filip and Ito,  Koichi},
  year = {2021},
  month = nov,
  pages = {104217}
}

@article{Zangen2016,
  title = {Are signalized intersections with cycle tracks safer? A case–control study based on automated surrogate safety analysis using video data},
  volume = {86},
  ISSN = {0001-4575},
  url = {http://dx.doi.org/10.1016/j.aap.2015.10.025},
  DOI = {10.1016/j.aap.2015.10.025},
  journal = {Accident Analysis \& Prevention},
  publisher = {Elsevier BV},
  author = {Zangenehpour,  Sohail and Strauss,  Jillian and Miranda-Moreno,  Luis F. and Saunier,  Nicolas},
  year = {2016},
  month = jan,
  pages = {161–172}
}

@article{althubaiti2016information,
  author  = {Althubaiti, Abdulrahman},
  title   = {Information bias in health research: definition, pitfalls, and adjustment methods},
  journal = {Journal of Multidisciplinary Healthcare},
  year    = {2016},
  volume  = {9},
  pages   = {211--217},
  doi     = {10.2147/JMDH.S104807},
  url     = {https://doi.org/10.2147/JMDH.S104807}
}

@article{ceccato2014safety,
  author  = {Ceccato, Vania},
  title   = {Safety on the move: Crime and perceived safety in transit environments},
  journal = {Security Journal},
  year    = {2014},
  volume  = {27},
  pages   = {127--131},
  doi     = {10.1057/sj.2014.11},
  url     = {https://doi.org/10.1057/sj.2014.11}
}

@article{marquezIntegratingPerceptionsSafety2021,
  title = {Integrating perceptions of safety and bicycle theft risk in the analysis of cycling infrastructure preferences},
  volume = {150},
  ISSN = {0965-8564},
  url = {http://dx.doi.org/10.1016/j.tra.2021.06.017},
  DOI = {10.1016/j.tra.2021.06.017},
  journal = {Transportation Research Part A: Policy and Practice},
  publisher = {Elsevier BV},
  author = {Márquez,  Luis and Soto,  Jose J.},
  year = {2021},
  month = aug,
  pages = {285–301}
}

@article{MANTON2016138,
  title = {Using mental mapping to unpack perceived cycling risk},
  volume = {88},
  ISSN = {0001-4575},
  url = {http://dx.doi.org/10.1016/j.aap.2015.12.017},
  DOI = {10.1016/j.aap.2015.12.017},
  journal = {Accident Analysis \& Prevention},
  publisher = {Elsevier BV},
  author = {Manton,  Richard and Rau,  Henrike and Fahy,  Frances and Sheahan,  Jerome and Clifford,  Eoghan},
  year = {2016},
  month = mar,
  pages = {138–149}
}

@inbook{parkinChapter6Network2012,
  title = {Chapter 6 Network Planning and Infrastructure Design},
  ISBN = {9781780522999},
  ISSN = {2044-9941},
  url = {http://dx.doi.org/10.1108/S2044-9941(2012)0000001008},
  DOI = {10.1108/s2044-9941(2012)0000001008},
  booktitle = {Cycling and Sustainability},
  publisher = {Emerald Group Publishing Limited},
  author = {Parkin,  John and Koorey,  Glen},
  year = {2012},
  month = may,
  pages = {131–160}
}

@article{parkinModelsPerceivedCycling2007,
  title = {Models of perceived cycling risk and route acceptability},
  volume = {39},
  ISSN = {0001-4575},
  url = {http://dx.doi.org/10.1016/j.aap.2006.08.007},
  DOI = {10.1016/j.aap.2006.08.007},
  number = {2},
  journal = {Accident Analysis \& Prevention},
  publisher = {Elsevier BV},
  author = {Parkin,  John and Wardman,  Mark and Page,  Matthew},
  year = {2007},
  month = mar,
  pages = {364–371}
}

@article{schrepp2017design,
  author  = {Schrepp, Martin and Hinderks, Andreas and Thomaschewski, J{\"o}rg},
  title   = {Design and Evaluation of a Short Version of the User Experience Questionnaire ({UEQ-S})},
  journal = {International Journal of Interactive Multimedia and Artificial Intelligence},
  year    = {2017},
  volume  = {4},
  number  = {6},
  pages   = {103--108},
  doi     = {10.9781/ijimai.2017.09.001},
  url     = {https://doi.org/10.9781/ijimai.2017.09.001}
}

@misc{ueqshort_items,
  author = {{User Experience Questionnaire}},
  title  = {Items of the Short Version of the User Experience Questionnaire (UEQ)},
  year   = {n.d.},
  url    = {https://www.ueq-online.org/Material/UEQS_Items.pdf},

}

@inproceedings{holmesChatbotUsabilityQuestionnaire,
  series = {ECCE 2019},
  title = {Usability testing of a healthcare chatbot: Can we use conventional methods to assess conversational user interfaces?},
  url = {http://dx.doi.org/10.1145/3335082.3335094},
  DOI = {10.1145/3335082.3335094},
  booktitle = {Proceedings of the 31st European Conference on Cognitive Ergonomics},
  publisher = {ACM},
  author = {Holmes,  Samuel and Moorhead,  Anne and Bond,  Raymond and Zheng,  Huiru and Coates,  Vivien and Mctear,  Michael},
  year = {2019},
  month = sep,
  pages = {207–214},
  collection = {ECCE 2019}
}

@article{zhangFeasibilityIntegratingWearable01,
  title = {The Feasibility of Integrating Wearable Cameras and Health Trackers for Measuring Personal Exposure to Urban Features: A Pilot Study in Roskilde,  Denmark},
  volume = {11},
  ISSN = {2160-9926},
  url = {http://dx.doi.org/10.4018/IJEPR.313181},
  DOI = {10.4018/ijepr.313181},
  number = {1},
  journal = {International Journal of E-Planning Research},
  publisher = {IGI Global},
  author = {Zhang,  Zhaoxi and Amegbor,  Prince Michael and Sabel,  Clive Eric},
  year = {2022},
  month = oct,
  pages = {1–21}
}

@inproceedings{jacobsenChatbotsDataCollection2025,
  series = {CHI ’25},
  title = {Chatbots for Data Collection in Surveys: A Comparison of Four Theory-Based Interview Probes},
  url = {http://dx.doi.org/10.1145/3706598.3714128},
  DOI = {10.1145/3706598.3714128},
  booktitle = {Proceedings of the 2025 CHI Conference on Human Factors in Computing Systems},
  publisher = {ACM},
  author = {Jacobsen,  Rune Møberg and Cox,  Samuel Rhys and Griggio,  Carla F. and van Berkel,  Niels},
  year = {2025},
  month = apr,
  pages = {1–21},
  collection = {CHI ’25}
}

@article{zhang2025human,
  title   = {Human-Chatbot Conversations About Urban Park Experiences: A Case Study from Washington Square Park in New York City},
  author  = {Zhang, Zhaoxi and Mendel, Tamir and Yin, Wen and Raipat, Vaidehi and Yabe, Takahiro},
  journal = {SSRN Electronic Journal},
  year    = {2025},
  month   = jan,
  doi     = {10.2139/ssrn.5378600},
  url     = {https://papers.ssrn.com/sol3/papers.cfm?abstract_id=5378600}
}

@Article{wevj16010020,
AUTHOR = {Yang, Boquan and Li, Jixiong and Zeng, Ting},
TITLE = {A Review of Environmental Perception Technology Based on Multi-Sensor Information Fusion in Autonomous Driving},
JOURNAL = {World Electric Vehicle Journal},
VOLUME = {16},
YEAR = {2025},
NUMBER = {1},
ARTICLE-NUMBER = {20},
URL = {https://www.mdpi.com/2032-6653/16/1/20},
ISSN = {2032-6653},
DOI = {10.3390/wevj16010020}
}

@article{VR2025Al,
  title = {Investigating the role of biophilic design to enhance comfort in residential spaces: human physiological response in immersive virtual environment},
  volume = {6},
  ISSN = {2673-4192},
  url = {http://dx.doi.org/10.3389/frvir.2025.1411425},
  DOI = {10.3389/frvir.2025.1411425},
  journal = {Frontiers in Virtual Reality},
  publisher = {Frontiers Media SA},
  author = {Al Sayyed,  Heba and Al-Azhari,  Wael},
  year = {2025},
  month = feb 
}

@misc{moran2025causal,
  doi = {10.48550/ARXIV.2507.04936},
  url = {https://arxiv.org/abs/2507.04936},
  author = {Moran,  Marcel and Salman,  Malik and Yabe,  Takahiro},
  title = {Causal Impacts of Protected Bike Lanes on Cycling Behavior with Demographic Disparities},
  publisher = {arXiv},
  year = {2025},
  copyright = {Creative Commons Attribution 4.0 International}
}

@article{itoUnderstandingUrbanPerception2024,
  title = {Understanding urban perception with visual data: A systematic review},
  volume = {152},
  ISSN = {0264-2751},
  url = {http://dx.doi.org/10.1016/j.cities.2024.105169},
  DOI = {10.1016/j.cities.2024.105169},
  journal = {Cities},
  publisher = {Elsevier BV},
  author = {Ito,  Koichi and Kang,  Yuhao and Zhang,  Ye and Zhang,  Fan and Biljecki,  Filip},
  year = {2024},
  month = sep,
  pages = {105169}
}

@article{Yu2022Exploring,
  title={Exploring built environment factors on e-bike travel behavior in urban China: A case study of Jinan},
  author={Yu, Yang and Jiang, Yang and Qiu, Ning and Guo, Huidong and Han, Xiaoran and Guo, Yanping},
  journal={Frontiers in Public Health},
  volume={10},
  pages={1013421},
  year={2022},
  publisher={Frontiers Media SA},
  doi={10.3389/fpubh.2022.1013421},
  url={https://doi.org/10.3389/fpubh.2022.1013421}
}

@article{Ewing01022009,
  title = {Measuring the Unmeasurable: Urban Design Qualities Related to Walkability},
  volume = {14},
  ISSN = {1469-9664},
  url = {http://dx.doi.org/10.1080/13574800802451155},
  DOI = {10.1080/13574800802451155},
  number = {1},
  journal = {Journal of Urban Design},
  publisher = {Informa UK Limited},
  author = {Ewing,  Reid and Handy,  Susan},
  year = {2009},
  month = feb,
  pages = {65–84}
}

@article{zhangUsingVirtualReality2026,
  title = {Using virtual reality to study human response to flood risk across controlled experiments},
  volume = {132},
  ISSN = {2212-4209},
  url = {http://dx.doi.org/10.1016/j.ijdrr.2025.105956},
  DOI = {10.1016/j.ijdrr.2025.105956},
  journal = {International Journal of Disaster Risk Reduction},
  publisher = {Elsevier BV},
  author = {Zhang,  Zhaoxi and Li,  Qinchan and Sun,  Qi and Ceferino,  Luis},
  year = {2026},
  month = jan,
  pages = {105956}
}

@article{falcoDigitalParticipatoryPlatforms2018,
  title = {Digital Participatory Platforms for Co-Production in Urban Development: A Systematic Review},
  volume = {7},
  ISSN = {2160-9926},
  url = {http://dx.doi.org/10.4018/IJEPR.2018070105},
  DOI = {10.4018/ijepr.2018070105},
  number = {3},
  journal = {International Journal of E-Planning Research},
  publisher = {IGI Global},
  author = {Falco,  Enzo and Kleinhans,  Reinout},
  year = {2018},
  month = jul,
  pages = {52–79}
}

@article{afzalanRoleSocialMedia2014,
  title = {The Role of Social Media in Green Infrastructure Planning: A Case Study of Neighborhood Participation in Park Siting},
  volume = {21},
  ISSN = {1466-1853},
  url = {http://dx.doi.org/10.1080/10630732.2014.940701},
  DOI = {10.1080/10630732.2014.940701},
  number = {3},
  journal = {Journal of Urban Technology},
  publisher = {Informa UK Limited},
  author = {Afzalan,  Nader and Muller,  Brian},
  year = {2014},
  month = jul,
  pages = {67–83}
}

@Article{land10010033,
AUTHOR = {Repette, Palmyra and Sabatini-Marques, Jamile and Yigitcanlar, Tan and Sell, Denilson and Costa, Eduardo},
TITLE = {The Evolution of City-as-a-Platform: Smart Urban Development Governance with Collective Knowledge-Based Platform Urbanism},
JOURNAL = {Land},
VOLUME = {10},
YEAR = {2021},
NUMBER = {1},
ARTICLE-NUMBER = {33},
URL = {https://www.mdpi.com/2073-445X/10/1/33},
ISSN = {2073-445X},
DOI = {10.3390/land10010033}
}

@article{raghunathanSplitQuestionnaireSurvey1995,
  title = {A Split Questionnaire Survey Design},
  volume = {90},
  ISSN = {1537-274X},
  url = {http://dx.doi.org/10.1080/01621459.1995.10476488},
  DOI = {10.1080/01621459.1995.10476488},
  number = {429},
  journal = {Journal of the American Statistical Association},
  publisher = {Informa UK Limited},
  author = {Raghunathan,  Trivellore E. and Grizzle,  James E.},
  year = {1995},
  month = mar,
  pages = {54–63}
}

@article{HE2026104115,
  title = {How a new metro line influences local traffic safety: A natural experiment in China},
  volume = {183},
  ISSN = {0967-070X},
  url = {http://dx.doi.org/10.1016/j.tranpol.2026.104115},
  DOI = {10.1016/j.tranpol.2026.104115},
  journal = {Transport Policy},
  publisher = {Elsevier BV},
  author = {He,  Dongsheng and An,  Zihao and Hu,  Yijia},
  year = {2026},
  month = jul,
  pages = {104115}
}

\end{document}